\documentclass[11pt]{article}

\usepackage{multirow}

\usepackage[nottoc]{tocbibind}

\usepackage{geometry}
\usepackage[affil-it, auth-sc]{authblk}
\usepackage[utf8]{inputenc}

\usepackage{graphicx}
\usepackage{amssymb}
\usepackage{amsthm}
\usepackage{amsmath}

\usepackage{color, soul}
\usepackage{verbatim}
\usepackage{lineno}
\usepackage{todonotes}
\presetkeys{todonotes}{inline, color=blue!30, size=\small}{}

\definecolor{darkred}{rgb}{0.9, 0.0, 0.0}
\definecolor{darkgreen}{rgb}{0.0, 0.5, 0.0}
\usepackage[colorlinks,bookmarks,linkcolor=darkred, citecolor=darkgreen]{hyperref}

\usepackage[ampersand]{easylist}

\usepackage{bm}
\usepackage{feynmf}

\newcommand{\be}{\begin{equation}}
\newcommand{\ee}{\end{equation}}
\newcommand{\bl}{\begin{easylist}}
\newcommand{\el}{\end{easylist}}

\newcommand{\LS}{\mbox{$\Lambda_{\rm singlet}$}}
\newcommand{\LT}{\mbox{$\Lambda_{\rm triplet}$}}

\newcommand{\FP}{\mbox{$F^{}_{P}$}}
\newcommand{\FA}{\mbox{$F^{}_{A}$}}
\newcommand{\FV}{\mbox{$F^{}_{1}$}}
\newcommand{\FM}{\mbox{$F^{}_{2}$}}
\newcommand{\ga}{\mbox{$g^{}_{A}$}}

\newcommand{\gaC}{\mbox{$\bar{g}^{}_{A}$}}
\newcommand{\gpC}{\mbox{$\bar{g}^{}_{P}$}}
\newcommand{\raq}{\mbox{$r^2_A$}}  

\newcommand{\FVCN}{\mbox{0.97578(8)}}
\newcommand{\FMN}{\mbox{3.70844}}
\newcommand{\FMCN}{\mbox{3.5986(82)}}
\newcommand{\gaN}{\mbox{1.2756(5)}}
\newcommand{\gaCN}{\mbox{1.2510(118)}}
\newcommand{\gpCN}{\mbox{8.25(25)}}
\newcommand{\gpMuCapN}{\mbox{8.23(83)}}
\newcommand{\raqN}{\mbox{0.46(22)}}
\newcommand{\raqAN}{\mbox{0.46(16)}}
\newcommand{\raqdipnuN}{\mbox{0.453(23)}}
\newcommand{\raqMuCapN}{\mbox{0.46(24)}}

\newcommand{\gpinn}{\mbox{$g_{\pi NN}$}}
\newcommand{\qq}{\mbox{$q^2$}}
\newcommand{\qqC}{\mbox{$q_0^2$}}
\newcommand{\hb}{$\chi$PT}

\newcommand{\fmq}{\mbox{fm$^2$}}
\newcommand{\si}{\mbox{s$^{-1}$}}
\newcommand{\ur}{\mbox{$u_r$}}

\newcommand{\ppm}{\mbox{$pp\mu$}}

\usepackage[sort&compress,numbers]{natbib}
\usepackage{doi}

\newcommand{\nl}{\nonumber \\ }
\newcommand{\order}{{\cal O}}

\allowdisplaybreaks

\begin{document}
%\linenumbers

\title{\Large\bf Nucleon Axial Radius and Muonic Hydrogen --\\
A New Analysis and Review}

\author[1,2,3]{Richard~J.~Hill}
\author[4]{Peter~Kammel}
\author[5]{William~J.~Marciano}
\author[6]{Alberto~Sirlin}
\affil[1]{Department of Physics and Astronomy, University of Kentucky, Lexington, KY 40506, USA  \vspace{1.2mm}}
\affil[2]{Fermilab, Batavia, IL 60510, USA  \vspace{1.2mm}}
\affil[3]{Perimeter Institute for Theoretical Physics, Waterloo, ON N2L 2Y5 Canada \vspace{1.2mm}}
\affil[4]{Center for Experimental Nuclear Physics and Astrophysics and Department of Physics, University of Washington, Seattle, WA 98195, USA
\vspace{1.2mm}}
\affil[5]{Department of Physics, Brookhaven National Laboratory, Upton, NY 11973, USA  \vspace{1.2mm}}
\affil[6]{Department of Physics, New York University, New York, NY 10003 USA  \vspace{1.2mm}} 

%\date{\today}                                           
\date{submitted April 19, 2018}                                           

\maketitle
\begin{abstract}  
Weak capture in muonic hydrogen ($\mu$H) as a probe of the chiral
properties and nucleon structure predictions of Quantum Chromodynamics
(QCD) is reviewed. A recent determination of the axial-vector charge
radius squared, $r_A^2(z\; {\rm exp.}) = \raqN\;\fmq$, from a
model independent $z$ expansion analysis of neutrino-nucleon
scattering data is employed in conjunction with the MuCap measurement
of the singlet muonic hydrogen capture rate,  $\Lambda_{\rm
  singlet}^{\rm MuCap} = 715.6(7.4)\;\si$, to update the
induced pseudoscalar nucleon coupling $\bar{g}_P^{\rm MuCap} =
\gpMuCapN$ derived from experiment, and  $\bar{g}_P^{\rm theory} =
\gpCN$ predicted by chiral perturbation theory.  Accounting for
correlated errors this implies $\bar{g}_P^{\rm theory}/\bar{g}_P^{\rm
  MuCap}= 1.00(8)$, confirming theory at the 8\% level.  If instead,
the predicted expression for $\bar{g}_P^{\rm theory}$ is employed as
input, then the capture rate alone determines $r_A^2(\mu {\rm
  H})=\raqMuCapN\, {\rm fm}^2$, or together with the independent $z$
expansion neutrino scattering result, a weighted average $r_A^2({\rm
  ave.}) = \raqAN\, {\rm fm}^2$.  Sources of theoretical uncertainty
are critically examined and potential experimental improvements are
described that can reduce the capture rate error by about a factor of
3.  Muonic hydrogen can thus provide  a precise and independent
$r_A^2$ value which may be compared with other determinations, such as
ongoing lattice gauge theory calculations.
The importance of an improved $r_A^2$ determination for phenomenology
is illustrated by considering the impact on critical neutrino-nucleus cross sections
at neutrino oscillation experiments.
\end{abstract}

\newpage

\tableofcontents

\newpage

\section{Introduction \label{sec:intro}}

Muonic hydrogen, the electromagnetic bound state of a muon and proton,
is a theoretically pristine atomic system.  As far as we know, it is
governed by the same interactions as ordinary hydrogen, but with the
electron of mass 0.511~MeV replaced by the heavier muon of mass
106~MeV, an example of electron-muon universality.
That mass enhancement ($\sim$207) manifests itself in much
larger atomic energy spacings and a smaller Bohr radius of $2.56\times
10^{-3}$\AA. This places the muonic hydrogen size about halfway (logarithmically) between the
atomic angstrom and the nuclear fermi (1 fm $= 10^{-5}$\AA) scale.

\begin{figure}[htb]
  \begin{center}
    \includegraphics[scale=1.25]{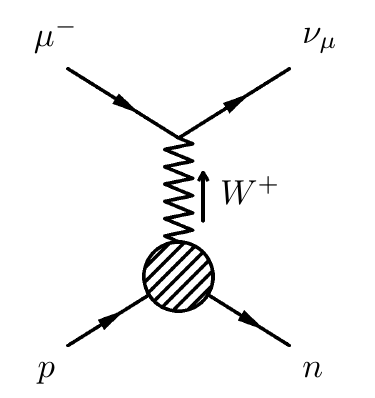}
    \caption{Muon capture on the proton, $\mu^-  p \to \nu_\mu  n$, via charged $W$ boson exchange. 
    \label{fig:capture}
    }
  \end{center}
\end{figure}

Those differences make muonic hydrogen very sensitive to otherwise
tiny effects such as those due to proton size and nucleon structure parameters
governing weak interaction phenomenology.  Indeed, muonic hydrogen
Lamb shift spectroscopy~\cite{Pohl:2010zza,Antognini:1900ns} has
provided a spectacularly improved measurement of the proton charge
radius that differs by about 7 standard deviations
from the previously accepted value
inferred from ordinary hydrogen and electron-proton scattering~\cite{Mohr:2015ccw}.
(The final verdict on this so called Proton Radius
Puzzle~\cite{Pohl:2013yb,Carlson:2015jba,Hill:2017wzi}
is still out.  For recent hydrogen spectroscopy measurements, see Ref.\cite{Beyer2017} and references therein).
Similarly, the larger muon mass kinematically allows the weak muon
capture process depicted in Fig.~\ref{fig:capture}, 
\begin{equation} \label{eq:omc}
\mu^{-} +  p \rightarrow \nu_\mu + n \,,
\end{equation}
to proceed, while ordinary hydrogen is (fortunately for our existence)
stable.  

Weak muon capture in nuclei has provided a historically important probe of
weak interactions and a window for studying nuclear structure. In
particular, weak capture in muonic hydrogen is a sensitive probe 
of the induced pseudoscalar component of the axial current
$p \to n$ matrix element which is well predicted from the chiral properties
of QCD. 
However, early experimental determinations of that pseudoscalar coupling,
\gpC,%
\footnote{The quantity \gpC\ is defined at the characteristic momentum $q_0^2$ for muon capture,
  see Eqs.~(\ref{eq:q02def}),(\ref{eq:gpCdef}) below.}
had, for some time, appeared problematic~\cite{Kammel:2010zz}.
All \gpC\ extractions from ordinary muon capture in hydrogen suffered
from limited precision, while the more sensitive extraction from
radiative muon capture~\cite{Wright:1998gi} disagreed with ordinary muon capture
and the solid prediction of
Chiral Perturbation
Theory (\hb)~\cite{Adler:1966gc,Wolfenstein:1970,Bernard:1994wn,Bernard:2000et,Kaiser:2003dr}.
An important underlying contribution to this problem was the chemical
activity of muonic hydrogen, which like its electronic sibling, can
form molecular ions, $(pp\mu)^+$. The highly spin dependent weak
interaction leads to very different capture rates from various muonic atomic and
molecular states. Thus, atomic physics processes like ortho-para
transitions in the muonic molecule, which flip the proton spins, significantly change
the observed weak capture rates and often clouded  the interpretation of
experimental results in the  55-year history of this field.
Unfortunately, the uncertainty induced by  molecular transitions was particularly
severe for the most precise measurements which were performed with high density
liquid hydrogen targets, where, because of rapid $pp\mu$ formation, essentially
capture from the molecule, not the $p\mu$ atom, is observed.
This problem was resolved by the MuCap Collaboration at the Paul Scherrer
Institute (PSI) which introduced an 
active, in situ, target, where ultra-pure hydrogen gas served both as the target as
well as the muon detector, thus enabling a measurement of the muonic hydrogen capture
rate at low density, where
\ppm\ formation is suppressed. MuCap unambiguously determined
the spin singlet muonic hydrogen capture rate $\Lambda_{\rm singlet}^{\rm MuCap} =
715.6(7.4) \,\si$~\cite{Andreev:2012fj, Andreev:2015evt} to 1\%
accuracy which,
when corrected for an enhancement from
radiative corrections~\cite{Czarnecki:2007th}, and  using
prevailing form factor values at the time implied $\bar{g}_P^{\rm MuCap}= 8.06(55)$, in excellent
agreement with $\bar{g}_P^{\rm theory} = 8.26(23)$, the predicted value.

We note, however, that the determination of $\bar{g}_P$ from
both experiment and theory required the input of the
axial charge radius squared,
traditionally taken from dipole form
factor fits to neutrino-nucleon quasielastic charged current scattering ($\nu_\mu n \to \mu p$) and pion
electroproduction ($e N \to e N^\prime \pi$) data,
which at the time implied the very precise $r_A^2(\rm dipole) =
0.454(13)\,\fmq$~\cite{Bodek:2007ym}.
Recently, that small ($\sim 3\%$) uncertainty in $r_A^2$ has been  called into question,
since it derives from the highly model dependent
dipole form factor assumption.%
\footnote{The dipole ansatz corresponds to $F_A(q^2) =
  F_A(0)/(1-q^2/m_A^2)^{2}$ with fit mass parameter $m_A$.}
The axial radius, which is central to this paper,  governs the momentum dependence of the axial-vector form factor,
by means of the expansion at small $q^2$,  
\begin{linenomath*}
\begin{align}\label{eq:FAlowq2}
F_A(q^2) = F_A(0) \left( 1 + \frac16 r_A^2 q^2 + \dots \right) \,. 
\end{align}
\end{linenomath*}
In the one-parameter dipole model, the terms denoted by the ellipsis
in Eq.~(\ref{eq:FAlowq2}) are completely specified in terms of $r_A^2$.
However, the true functional form of $F_A(q^2)$ is unknown, and the
dipole constraint represents an uncontrolled systematic error. 
We may instead employ the $z$ expansion formalism, a convenient method for 
enforcing the known complex-analytic structure of the form factor inherited from QCD, while avoiding
poorly controlled model assumptions.
This method replaces the dipole $F_A(q^2)$ with  $F_A[z(q^2)]$, which in terms of the conformal mapping variable $z(q^2)$,
has a convergent Taylor expansion for all spacelike $q^2$.  The size of the expansion
parameter, and the truncation order of the expansion
necessary to describe data of a given precision in a specified
kinematic range, are determined a priori. 
This representation helps ensure that observables extracted from data are not influenced by implicit form
factor shape assumptions. 
Using the $z$ expansion~\cite{Bhattacharya:2011ah} to fit the
neutrino data alone leads to~\cite{Meyer:2016oeg} $r_A^2(z\; {\rm exp.},\; \nu) = \raqN\,\fmq$
with a larger ($\sim 50\%$), more
conservative but better justified error.
As we will discuss below, traditional analyses of pion electroproduction data have also used a
dipole assumption to extract $r_A^2$ from $F_A(q^2)$, and in addition
required the a priori step of
phenomenological modeling to extract $F_A(q^2)$ from data.  Since these model
uncertainties have not been quantified, we refrain from including pion electroproduction
determinations of $r_A^2$ in our analysis.  Similarly, we do not include extractions
from neutrino-nucleus scattering on nuclei larger than the deuteron, in order
to avoid poorly quantified nuclear model uncertainties.
In this context,
we note that dipole fits to recent $\nu$-C scattering data suggest a smaller 
$\raq \approx 0.26\, \mbox{fm}^2$~\cite{AguilarArevalo:2010zc},
compared to
historical dipole values $\raq \sim 0.45\, \mbox{fm}^2$~\cite{Bodek:2007ym}.  
This discrepancy may be due to form factor shape biases~\cite{Bhattacharya:2011ah}
(i.e., the dipole assumption),
mismodeling of nuclear effects~\cite{Alvarez-Ruso:2017oui,Benhar:2005dj,Ankowski:2005wi,Ankowski:2010yh,Juszczak:2010ve,Amaro:2010sd,Nieves:2011yp,Martini:2012fa,Nieves:2012yz}, or something else. 
Independent determination of $r_A^2$ is a necessary ingredient for resolving
this discrepancy.
Finally, we do not include recent interesting lattice QCD results~\cite{Green:2017keo,Alexandrou:2017hac,Capitani:2017qpc,Gupta2017,Yao2017}, 
some of which suggest considerably smaller $r_A^2$ values.
As we shall discuss
below in Sec.~\ref{sec:impact}, future improvements on these lattice QCD
results could provide an independent $r_A^2$ value with controlled
systematics, that would open new opportunities for interpreting muon capture.
To illustrate the broad range of possible $r_A^2$ values, we provide in Table~\ref{tab:rA2vals}
some representative values considered in the recent literature.

\begin{table}[htp]
  \caption{Illustrative values obtained for $r_A^2$ from neutrino-deuteron quasi elastic scattering ($\nu$-$d$),
    pion electroproduction ($eN \to eN^\prime \pi$), neutrino-carbon quasielastic scattering ($\nu$-C),
    muon capture (MuCap) and lattice QCD.
    Values labeled ``dipole” enforce the dipole shape ansatz.
    The value labeled ``$z$\;exp.'' uses the model independent $z$ expansion. 
  }
\begin{center}
\begin{tabular}{rclc}
  \hline\hline
\\[-3mm]
Description\quad	&&$r_A^2\;({\rm fm}^2)$  &Source/Reference\vspace{1mm}\\ 
\hline\hline
\\[-3mm]
$\nu d$ (dipole) && $0.453(23)$ & \cite{Bodek:2007ym}
\\
$e N \to e N^\prime \pi$ (dipole) && $0.454(14)$ & \cite{Bodek:2007ym}
\\
\hline
average\;\, && $0.454(13)$
\\[1mm]
\hline\hline
\\[-3mm]
$\nu$C (dipole) && $0.26(7)$ & \cite{AguilarArevalo:2010zc}  
\\[1mm]
\hline\hline
\\[-3mm]
$\nu d$ ($z$\;exp.) && $0.46(22)$ & \cite{Meyer:2016oeg}
\\
MuCap\;\,  && \raqMuCapN & this work
\\
\hline
average\;\, && \raqAN
\\[1mm]
\hline\hline
\\[-3mm]
\multirow{3}{*}{lattice QCD\;\,}
&& $0.213(6)(13)(3)(0)$ & \cite{Green:2017keo}
\\
&& $0.266(17)(7)$ & \cite{Alexandrou:2017hac}
\\
&& $0.360(36)^{+80}_{-88}$ & \cite{Capitani:2017qpc}
\\
&& $0.24(6)$ & \cite{Gupta2017}
\\[1mm]
\hline\hline
\end{tabular}
\end{center}
\label{tab:rA2vals}
\end{table}

Accepting the larger $r_A^2$
uncertainty from the $z$ expansion fit to neutrino data, leads to renewed
thinking about the utility of precision
measurements of muonic hydrogen capture rates for probing QCD chiral
properties. As we shall see, the determination of \gpC\ becomes
$\bar{g}_P^{\rm MuCap}=\gpMuCapN$ and $\bar{g}_P^{\rm theory}=\gpCN$, which
are still in good agreement, but with errors enlarged by factors of 1.7 and
3.5, respectively, compared to results using $r_A^2(\rm dipole)$ [cf. Eqs.~(\ref{eq:gp_dipole}),(\ref{eq:gp_z}) below].  However, taking into account the correlated uncertainties,
the comparison can be sharpened to $\bar{g}_P^{\rm theory}/\bar{g}_P^{\rm MuCap} =1.00(8)$.

Instead of determining \gpC, one can use the
functional dependence of this quantity, 
$\gpC(r_A^2)$, 
predicted from \hb\ to extract $r_A^2$
from the singlet capture rate.  As we shall show,
that prescription
currently gives a sensitivity to $r_A^2$ comparable to $z$-expansion
fits to neutrino-nucleon scattering.  We use the resulting value from muon
capture to derive a combined weighted average.   We  also examine how such a method
can be further improved by better theory and experiment, and
demonstrate that a factor of $\sim$3 improvement in the experimental
precision appears feasible and commensurate with our updated
theoretical precision.

The axial radius is indispensable for ab-initio calculations of
nucleon-level charged current quasielastic cross sections
needed for the interpretation of long baseline neutrino oscillation
experiments at $|q^2| \sim 1\, {\rm GeV}^2$.  Its current uncertainty is
a serious impediment to
the extraction of neutrino properties from such
measurements.  We quantify the impact that an improved muon capture
determination of $r_A^2$ would have on neutrino-nucleon cross
sections, and discuss the status and potential for other
determinations, particularly from the promising lattice QCD approach.

The remainder of this paper is organized as follows: In
Sec.~\ref{sec:theory} we give an overview and update regarding the
theory of $\mu$-$p$ capture in muonic hydrogen. 
After describing the lowest order formalism, we discuss the magnitude and uncertainty of radiative corrections
(RC) to muon capture. Normalizing
relative to superallowed and neutron beta decays, we argue that the
uncertainty in the singlet muon capture rate from radiative corrections is
much smaller than the conservative estimate of $\pm$0.4\% originally given in Ref.~\cite{Czarnecki:2007th}. 
Based on further considerations, we now estimate it to be $\pm$0.1\%.
 Uncertainties in the input parameters are described,
with particular emphasis on a numerical analysis of the axial charge
radius squared and its potential extraction from the singlet 1S
capture rate in $\mu H$.  Then, in Sec.~\ref{sec:experiment}, we describe the
experimental situation.  After reviewing the MuCap result, we discuss
possible improvements for a next generation experiment that would aim
for a further factor of $\sim$3 error reduction. In Sec.~\ref{sec:reach}, we
discuss what can be learned from the present MuCap result and an
improved experiment.
We update the determination of \gpC\ from the MuCap measurement using the more
conservative $z$ expansion value of $r_A^2$ obtained from neutrino-nucleon scattering.
Then, as a change in strategy, using the theoretical expression for
\gpC\ obtained from \hb\ as input,
$r_A^2$ is extracted from the MuCap capture rate and averaged with the $z$ expansion
value. Other utilizations of MuCap results are also discussed.
In Sec.~\ref{sec:impact}, we illustrate the impact of an improved $r_A^2$
determination
on quasielastic neutrino scattering cross sections and discuss the status
of, and prospects for, improving  alternative $r_A^2$ determinations.
Section~\ref{sec:discussion} concludes with a
summary of our results and an
outlook for the future.

\section{Muon capture theory update}
\label{sec:theory}

The weak capture process, Eq.~\eqref{eq:omc}, from a muonic hydrogen bound state is a multi-scale
field theory calculational problem, involving electroweak, hadronic and atomic mass scales.  In
this section, we review the essential ingredients of this problem before
discussing the status of phenomenological inputs and the numerical evaluation
of the capture rate $\Lambda$. 
The calculation can be viewed as an expansion in small parameters, $\alpha \sim m_\mu^2/m_p^2 \sim \epsilon^2$,
and will result in a structure
\be\label{eq:expansion}
\Lambda \sim \bigg[ F_1, F_A \bigg] + \epsilon \bigg[ F_2, \gpC \bigg]
+ \epsilon^2 \bigg[ r_1^2, r_A^2 \bigg]
+ \order(\epsilon^3) \,, 
\ee
where in this formula $F_i$ denotes a form factor at $q^2=0$,  $r_i^2$ is the corresponding radius defined
as in Eq.~(\ref{eq:FAlowq2}) and \gpC\ the
pseudoscalar coupling at \qqC.
Thus to achieve permille accuracy on the capture rate (where $10^{-3} \sim \epsilon^3$),
we require all $\order(\epsilon^2)$ corrections with $\pm10\%$ precision, and $\order(\epsilon^3)$ corrections with
$\pm 100\%$ precision.

\subsection{Preliminaries} 

\par

For processes at low energy, $E \ll m_W$, where $m_W \approx 80\,{\rm GeV}$ is the
weak charged vector boson mass, the influence of heavy particles and other physics at the weak scale
is rigorously encoded in the parameters of an effective Lagrangian
containing four-fermion operators.
For muon capture the relevant effective Lagrangian is 
\begin{linenomath*}
\begin{align}\label{eq:eff_op}
  {\cal L} = - {G_F V_{ud}  \over \sqrt{2}}
  \bar{\nu}_\mu \gamma^\mu (1-\gamma^5) \mu \;\bar{d} \gamma_\mu (1-\gamma^5) u  + {\rm H.c.} + \dots \,,
\end{align}
\end{linenomath*}
where $G_F$ and $V_{ud}$ are  the Fermi constant and the CKM  up-down quark mixing parameter
respectively (cf. Table~\ref{tab:input}),
and the ellipsis denotes effects of radiative corrections.
Atomic physics of the muonic hydrogen system is described by the
effective Hamiltonian, valid for momenta satisfying $|\bm{p}| \ll m_\mu$: 
\begin{linenomath*}
\begin{align}\label{eq:H}
  H = {p^2\over 2 m_r} -{\alpha \over r} + \delta V_{\rm VP}
  - i {G_F^2|V_{ud}|^2 \over 2} \left[ c_0 + c_1 ( \bm{s}_\mu + \bm{s}_p)^2 \right] \delta^3(\bm{r}) \,,
\end{align}
\end{linenomath*}
where $m_r=m_\mu m_p /(m_\mu + m_p)$ is the reduced mass,
$\delta V_{\rm VP}$ accounts for electron vacuum polarization as discussed below,
and $\bm{s}_\mu$, $\bm{s}_p$ are muon and proton spins.  
The annihilation process is described by an anti-Hermitian component of $H$~\cite{Hill:2000qi}.
Since the weak annihilation is a short-distance process compared to atomic length scales,
this anti-Hermitian component can be expanded as a series of local operators. At
the current level of precision terms beyond the leading one, $\delta^3(\bm{r})$, are irrelevant~\cite{Hill:2000qi}. 
Relativistic corrections to the Coulomb interaction in Eq.~(\ref{eq:H})
are similarly irrelevant~\cite{Lepage:1997cs}.  In both cases, neglected operators
contribute at relative order $v^2/c^2 \sim \alpha^2$, where $v$ is the nonrelativistic
bound state velocity. 
Electron vacuum polarization enters formally at order $\alpha^2$, but is enhanced by a factor
$m_\mu/m_e$ making it effectively a first order correction~\cite{Eides:2000xc,Eiras:2000rh}.  

Having determined the structure of the effective Hamiltonian (\ref{eq:H}), 
the coefficients $c_i$ are determined by a matching condition with the quark level theory (\ref{eq:eff_op}). 
The annihilation rate in the 1S state is then computed from $H$ to be 
\begin{linenomath*}
\begin{align}\label{eq:rate}
  \Lambda &= {G_F^2|V_{ud}|^2} \times  \left[  c_0 + c_1 F(F+1) \right] \times |\psi_{\rm 1S}(0)|^2 + \dots 
  \,, 
\end{align}
\end{linenomath*}
where $|\psi_{\rm 1S}(0)|^2 = m_r^3\alpha^3/\pi$ is the ground state
wavefunction at the origin squared and $F$ is the total spin ($F=0$ for singlet, $F=1$ for triplet).  
Equation~(\ref{eq:rate}), with $c_i$ expressed in terms of hadronic form factors
(cf. Eq.~(\ref{eq:ci}) below), 
exhibits the factorization of the process into contributions arising  from
weak, hadronic and atomic scales. 

\subsection{Tree level calculation}

Hadronic physics in the nucleon matrix elements of the vector and axial-vector quark currents
of Eq.~(\ref{eq:eff_op}) is parameterized as:%
\footnote{We choose a convention for the pseudoscalar form factor that is
  independent of lepton mass: $F_P(q^2)=(m_N/m_\mu)g_P(q^2)$, in terms of $g_P(q^2)$ used
  in Ref.~\cite{Andreev:2007wg}.  Our sign conventions for $F_A$ and $F_P$ are such that
  $F_A(0)$ and $F_P(0)$ are positive.}
\begin{linenomath*}
\begin{align} \label{eq:VA}
\langle n|(V^\mu-A^\mu) |p \rangle &=\bar{u}_n\!\bigg[ \FV(q^2)\gamma^\mu
  \!+\frac{i \FM(q^2)}{2m_N}\sigma^{\mu\nu}q_\nu
  - \text{\FA}(q^2)\gamma^\mu\gamma^5
  -\frac{ \text{\FP}(q^2)}{m_N}q^\mu\gamma^5
  \nl
  & \qquad
  + \frac{ F_S(q^2) }{ m_N } q^\mu
  - \frac{ i F_T(q^2) }{2 m_N} \sigma^{\mu\nu} q_\nu \gamma^5
  \bigg]
u_p + \dots \,,
\end{align}
\end{linenomath*}
where $V^\mu - A^\mu = \bar{d}\gamma^\mu u - \bar{d}\gamma^\mu \gamma^5 u$, 
and the ellipsis again denotes effects of radiative corrections.  
For definiteness we employ the average nucleon mass $m_N \equiv (m_n+m_p)/2$.  
The form factors $F_S$ and $F_T$ are so-called second class amplitudes that 
violate $G$ parity and are suppressed by isospin violating quark masses
or electromagnetic couplings~\cite{Shiomi:1996np,Govaerts:2000ps,Minamisono:2001cd,Holstein:2014fha}.
They would appear in
the capture rate, Eq.~(\ref{eq:ci}) below,
accompanied by an additional factor $m_\mu/m_N$ relative to $F_1$ and $F_A$.
Similar to isospin violating effects in $F_2(0)$, discussed below in Sec.~\ref{sec:inputs}, 
power counting predicts negligible impact of $F_S$ and $F_T$ at the permille level; we thus ignore them in
the following discussion. 

The $c_i$ in Eq.~(\ref{eq:rate}) are determined by matching
the quark level theory (\ref{eq:eff_op}) to the nucleon level theory (\ref{eq:H}),
using the hadronic matrix elements (\ref{eq:VA}).  
This matching is accomplished by enforcing, e.g., equality of 
the annihilation rate for $\mu  p \to \nu_\mu  n$ computed in both theories for the limit of
free particles, with the proton and muon at rest. 
For the coefficients corresponding to singlet and triplet decay rates,
this yields~\cite{Santisteban1976,Czarnecki:2007th} 
\begin{linenomath*}
\begin{align}\label{eq:ci}
  c_0 &=  {E_\nu^2\over 2 \pi M^2} (M-m_n)^2
  \bigg[ {2M-m_n \over M-m_n} \FV(q_0^2) + {2M + m_n \over M-m_n} \FA(q_0^2) - {m_\mu \over 2 m_N} \FP(q_0^2)
  \nl
  &\quad 
  + (2M + 2m_n - 3 m_\mu){\FM(q_0^2) \over 4 m_N}
  \bigg]^2 
  \,,
  \nl
  c_0+ 2 c_1 & = {E_\nu^2\over 24 \pi M^2} (M-m_n)^2
  \bigg\{ \bigg[ {m_\mu \over m_N} F_P(q_0^2) - {2m_n \over M-m_n} \big(F_{1}(q_0^2)-F_A(q_0^2) \big)
  \nl
  &\hspace{-5mm}
    + (2M + 2m_n - m_\mu) {\FM(q_0^2)\over 2m_N} \bigg]^2
  + 2\bigg[ {m_\mu \over m_N} \FP(q_0^2) + {2M \over M- m_n} \big(\FV(q_0^2)-F_A(q_0^2)\big) - m_\mu {\FM(q_0^2) \over 2 m_N} \bigg]^2 \bigg\} 
  \,, 
\end{align}
\end{linenomath*}
where the initial state mass is $M \equiv m_\mu + m_p$, the neutrino
energy is $E_\nu \equiv (M^2-m_n^2)/2M = 99.1482~\mbox{MeV}$, and
the invariant momentum transfer is
\begin{linenomath*} 
\begin{align}\label{eq:q02def}
  q_0^2  \equiv
  m_\mu^2 - 2 m_\mu E_\nu 
  =   -0.8768\, m_\mu^2.
\end{align}
\end{linenomath*}
Since the matching is performed with free particle states, the 
quantities $M$, $E_\nu$ and $q_0^2$ are defined independent of the atomic binding
energy, as necessary for determination of the state-independent coefficients $c_i$ of
the effective Hamiltonian (\ref{eq:H}).%
\footnote{In particular, a binding energy is not included in the initial-state mass $M$, but
  would anyways correspond to a relative order $\alpha^2$ correction that is beyond the current
  level of precision.}

The amplitudes (\ref{eq:ci}) can also be expressed as an expansion in 
\hb~\cite{Bernard:2000et,Raha:2013lqa,Pastore:2013nsa,Pastore:2014iga}.
However, the general formulas in Eq.~(\ref{eq:ci}) allow us to more directly
implement and interpret experimental constraints on the form factors and do not carry the
intrinsic truncation error of NNLO \hb\ derivations (estimated in Ref.~\cite{Pastore:2014iga} as $\pm 1\%$).
For example, we may take the vector form factors $F_{1}$, $F_{2}$ directly from
experimental data, rather than attempting to compute them as part of an expansion in \hb. 
%
%% No approximation is yet made in Eq.~\eqref{eq:ci}, except for
%% neglect of second class currents, as justified above, and neglect of contributions on the left hand
%% side of Eq.~(\ref{eq:ci}) at relative order $\alpha^2$, as discussed after Eq.~(\ref{eq:H}).
%
We investigate below the restricted application of \hb\ to express $F_P(q_0^2)$ 
in terms of $r_A^2$ and other experimentally measured quantities. 

\subsection{Radiative corrections}

\begin{figure}[tb]
  \begin{center}
    \includegraphics[scale=1.25]{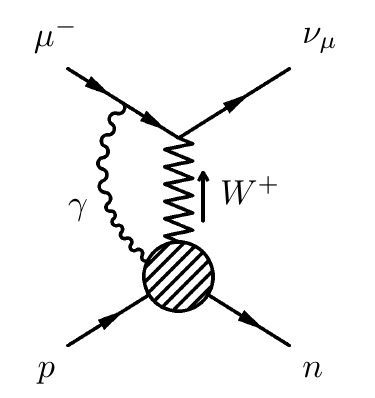}
    \caption{Example of an $\order(\alpha)$ $\gamma W$ exchange
      box diagram radiative correction to muon capture.  
    \label{fig:radcor}
    }
  \end{center}
\end{figure}

The electroweak radiative corrections to muon capture in muonic
hydrogen, depicted in Fig.~\ref{fig:radcor}, were first
calculated in Ref.~\cite{Czarnecki:2007th}.  Here,
we briefly describe the origin of such quantum loop effects and take
this opportunity to update and reduce their estimated uncertainty.  The
computational strategy relies on the well known electroweak
corrections to (i) the muon lifetime~\cite{Sirlin:1977sv,Tishchenko:2012ie}, (ii) super-allowed $0^+ \to
0^+$ $\beta$ decays~\cite{Sirlin:1977sv,Hardy:2014qxa,Marciano:2005ec}, and (iii) the neutron lifetime~\cite{Czarnecki:2004cw,Patrignani:2016xqp,Czarnecki2018}.

Radiative corrections to weak decay processes in the Standard Model involve
ultraviolet divergences that can be renormalized, 
yielding finite phenomenological parameters such as the Fermi
constant $G_F$  obtained from the measured muon lifetime~\cite{Tishchenko:2012ie} and the CKM
matrix element $|V_{ud}|$  obtained from super-allowed $\beta$ decays
(see Table~\ref{tab:input}). 
In terms of those parameters, the radiative corrections to the
neutron lifetime and the muon capture rate are rendered finite
and calculable.
We note that the matrix element of the vector current is absolutely normalized
at $q^\mu=0$, corresponding to a Conserved Vector Current (CVC): $F_1(0)=1$,
up to second order corrections in small isospin violating
parameters~\cite{Behrends:1960nf,Terentev1963,Ademollo:1964sr}. 
On the other hand, the normalization of the remaining form factors
appearing in Eq.~(\ref{eq:ci}) requires a conventional definition in 
the presence of radiative corrections.  This definition is specified
at $q^2=0$ by a factorization requirement that 
expresses the total process as a tree level expression times
an overall radiative correction.
For example, the neutron decay rate in this scheme involves the factor 
$(1+3 g_A^2)(1+{\rm RC})$, where $(1+3 g_A^2)$ is the tree level expression
with $F_A(0)=g_A$, and RC denotes the radiative corrections.
With that definition of \ga, the factorized radiative corrections are taken to be the same for vector and axial-vector 
contributions but actually computed for the vector part which is cleaner theoretically because of CVC. 
Radiative corrections that are different for the axial current amplitude are then, by definition, absorbed into \ga.
That well defined \ga\ can be precisely obtained from the neutron lifetime, $\tau_n$, in conjunction with $V_{ud}$ determined from superallowed beta decays via the relationship~\cite{Marciano:2005ec,Czarnecki:2004cw, Patrignani:2016xqp,Czarnecki2018}

\begin{align}\label{eq:gA_Vud_taun}
\left( 1+3 g_A^2 \right)  |V_{ud}|^2 \tau_n = 4908.6(1.9)\; {\rm s} \,,
\end{align}
where the uncertainty comes primarily from radiative corrections to
the neutron lifetime.
Alternatively, that same \ga\ can be directly obtained from neutron decay final state asymmetries after accounting for small QED corrections~\cite{Shann:1971fz}
and residual Coulomb, recoil and weak magnetism effects~\cite{Wilkinson:1982hu}.
We employ the lifetime method here, because it is currently more precise.

In the case of muon capture, we have four form factors all
evaluated at $q_0^2$: vector ($F_1$), induced weak magnetism
($F_2$), axial-vector ($F_A$) 
and induced pseudoscalar ($F_P$). We define these form factors to all
have the same electroweak radiative corrections and explicitly compute
those corrections for $F_1$. 

Short-distance corrections (which dominate) correspond to 
a renormalization of the relevant quark-level four-fermion operator, and 
are automatically the same for all form factors.
Long distance corrections are
smaller but significant. They include contributions from the $g$-function computed for neutron decay~\cite{Sirlin:1967zza}  which contains lepton energy
dependent corrections not affected
by strong interactions that are the same for the vector and axial-vector form factors.
The $g$-function is also the same for both neutron decay and muon capture but differs significantly in magnitude for the two cases because of the
different charged lepton masses and kinematics involved in the two processes.
In addition to the $g$-function, there are strong interaction dependent
constant vector and axial-vector radiative corrections that can be estimated (for the vector amplitude) and factored into the RC. Differences
in the axial amplitude can then be absorbed into a redefinition of \ga.
Invoking G-parity and charge-symmetry of the strong interactions, they can be shown to be the same in neutron decay and
muon capture, up to small O($\alpha \, m_{\mu}/m_p$)
corrections. Hence, the \ga\ definition is essentially the same for both processes.
That feature is important for reducing the radiative correction uncertainty
in muon capture.

\begin{comment}
In terms of the above parameters ($G_F$, $V_{ud}$ and \ga),  the radiative
corrections to the neutron decay rate are finite and known; thus they
can be used n conjunction with $G_F$ and $V_{ud}$ to
determine \ga\ from the neutron lifetime.  
For the case of muon capture,
we have four contributions to the amplitude: vector ($F_1$), induced weak
magnetism ($F_2$), axial-vector ($F_A$) and induced pseudo scalar ($F_P$).
We define those form factors such that the radiative corrections are
same for all of them, as explicitly computed for the vector.
\todo{Bill slightly modified, please check}
That assumption can be taken to define the form factor normalizations
at the momentum transfer \qqC\ for muon capture.%
\footnote{Note that any discrepancy between the radiative corrections to
  different form factors in an alternate scheme is by definition absorbed into the form
  factors, and should be accounted for when the form factors are
  determined by other processes.}

Having specified the definition of the form factors
appearing in muon capture, we need only compute the radiative corrections to the
vector amplitude and include them as an overall factor in the rate.
\end{comment}
Modulo the radiative corrections absorbed into the form
factor definitions, the 1+RC common radiative correction factor applicable
to muon capture can be written as the sum of three terms
\begin{linenomath*}
\begin{align}\label{eq:RCtot}
  {\rm RC} &= {\rm RC}(\rm electroweak) + {\rm RC}(finite\,\,size) + {\rm RC}(electron\,\,VP)   \,,
\end{align} 
\end{linenomath*}
which we now specify. 
Neglecting terms of $\order(E_\ell/m_p, q/m_p)$, where $E_\ell$
is the charged lepton energy and $q$ the momentum transfer,%
\footnote{For the kinematics of muon capture, $E_\ell/m_p \sim q/m_p \sim m_\mu/m_p$.}
the radiative corrections to the vector parts of
neutron decay and muon capture are of the same form, but evaluated at
different $q^2$ and with 
different lepton mass. 
The RC (electroweak) radiative corrections to muon
capture~\cite{Czarnecki:2007th} were obtained from the
original neutron decay calculation, but 
including higher-order leading log effects
denoted by ellipsis in the following Eq.~(\ref{eq:ew_radcor}): 
\begin{linenomath*} 
\begin{align}
  \label{eq:ew_radcor}
  {\rm RC}(\rm electroweak) &= {\alpha\over 2\pi}
  \bigg[ 4 \ln{m_Z \over m_p} -0.595 + 2C + g(m_\mu, \beta_\mu=0) \bigg] + \dots 
= + 0.0237(4)
  \,,
\end{align}
\end{linenomath*}
where $m_Z =91.1876\,{\rm GeV}$, $m_p=0.9383\,{\rm GeV}$, $C=0.829$~\cite{Marciano:2005ec}, and
$g(m_\mu, \beta_\mu=0) = 3\log(m_p/ m_\mu) -  27/4 = -0.199$
was obtained from Eq.~(20b) in Ref.~\cite{Sirlin:1967zza}
by replacing $m_e \to m_\mu$, ignoring bremsstrahlung and taking the
$\beta_\mu =0$ limit. 
The higher order (in $\alpha$) corrections enhanced by large
logarithms~\cite{Czarnecki:2004cw}
have been added to the $+2.23\%$ order $\alpha$ correction
to obtain the 
total $+2.37\%$ electroweak radiative correction.

In Eq.\eqref{eq:ew_radcor} the uncertainty in the electroweak
radiative corrections has been reduced  by an order of magnitude to about
$\pm 4 \times 10^{-4}$ compared with the very conservative
$\pm 4 \times 10^{-3}$ value in Ref.~\cite{Czarnecki:2007th}.
The source of this uncertainty is the axial-vector induced
-0.595 + 2C terms in Eq.\eqref{eq:ew_radcor} contributing to the RC in
fig.~\ref{fig:radcor}.  In Ref.~\cite{Marciano:2005ec}
an assumed 10\% uncertainty in C and 100\% uncertainty in the matching of long
and short distance radiative corrections were evaluated to contribute $\pm 2\times 10^{-4}$
and $\pm 3\times 10^{-4}$, respectively and an error of $\pm 1\times10^{-4}$ was assigned to possible unaccounted 
for 2 loop effects. Taken together in quadrature they give an overall error of $\pm 4\times 10^{-4}$ 
for the electroweak radiative corrections to muon capture. That estimate is essentially the same as the error found for 
the universal radiative corrections to neutron decay and superallowed beta decays~\cite{Marciano:2005ec}.
As a result, the $\pm 4\times 10^{-4}$ error manifests itself also in
$|V_{ud}|^2$ when extracted from superallowed beta decays where it is anti-correlated with the corresponding errors in 
neutron decay and muon capture. Consequently, when $|V_{ud}|^2$
obtained from superallowed beta decays is used in 
neutron decay and muon capture calculations, the $\pm 4 \times 10^{-4}$ uncertainties
cancel and one is left with essentially no electroweak RC uncertainty
at least to O($10^{-4}$).
Indeed, even if we were more conservative by increasing these
uncertainties by a common factor, they would continue to
cancel. That cancellation and the common \ga\ definition in muon capture and neutron decay together help to 
keep the electroweak RC uncertainty small.
There are, however, contributions to muon capture of  O($\alpha\, m_{\mu}/m_p$) in the box diagram of
Fig.~\ref{fig:radcor} that could potentially be significant. However, a recent 
calculation\footnote{A. Sirlin, unpublished}, that ignores
nucleon structure, found such contributions to either cancel among themselves or have small coefficients that render them negligible.

Next, we assume that corrections of $\order(\alpha\, m_\mu/m_p)$ due to nucleon structure
are parametrized by the nucleon finite size reduction factor~\cite{Friar:1978wv}
\begin{linenomath*}
\begin{align}\label{eq:FSansatz}
|\psi_{\rm 1S}(0)|^2 \to {m_r^3\alpha^3 \over \pi}\left( 1 - 2 \,\alpha\, m _r \langle r \rangle \right) \,,
\end{align}
\end{linenomath*}
where $\langle r \rangle$ denotes the first moment of the proton charge
distribution. 
Based on a range of model forms for this distribution, the correction (\ref{eq:FSansatz})
evaluates to 
\begin{linenomath*}
\begin{align}\label{eq:FS} 
{\rm RC}(\rm finite\,\,size) &= -0.005(1) \,,
\end{align}
\end{linenomath*}
where the error, which is the current dominant overall uncertainty in the RC, spans the central values $-0.0044$~\cite{Govaerts:2000ps}, 
$-0.005$~\cite{Czarnecki:2007th},
and $-0.0055$~\cite{Raha:2013lqa} given in the literature. 
We note that the quoted uncertainty may not fully account
for possible additional effects of nucleon structure which could be estimated
using a relativistic evaluation of the $\gamma$-$W$ box diagrams
including structure dependence,
but are beyond the scope of this article.%  
\footnote{The finite size ansatz (\ref{eq:FSansatz})
  becomes exact in the large-nucleus limit, $r_{\rm nucleus} \gg r_{\rm weak}$,
  where
  $r_{\rm nucleus} \sim r_{E,p}$
  is the nuclear (proton) charge radius and $r_{\rm weak} \in (r_1,\, r_2,\, r_A)$
  denotes a weak vector or axial radius.}

The corrections RC(electroweak) and RC(finite size) modify the coefficients $c_i$
of the effective Hamiltonian (\ref{eq:H}).  The remaining radiative correction,
from the electron vacuum polarization modification to
the muonic atom Coulomb potential, is described by $\delta V_{\rm VP}$. 
This contribution amounts to
\begin{linenomath*}
\begin{align}
{\rm RC}(\rm electron\,\,VP) &=  +0.0040(2), 
\end{align}
\end{linenomath*}
where the very small uncertainty 0.02\% is estimated by the difference between
$1.73\,\alpha/\pi$ of Ref.~\cite{Czarnecki:2007th,Eides:2000xc} and
$1.654\,\alpha/\pi$ of Ref.~\cite{Raha:2013lqa}. 

In Eq.~(\ref{eq:RCtot}), we have defined the total radiative correction to
include electroweak, finite size and electron vacuum polarization contributions.
In Ref.~\cite{Czarnecki:2007th}, the finite size correction was treated separately,
and ``radiative correction'' referred to the sum of our RC(electroweak) and RC(electron VP),
amounting to 2.77\%.  That central value of the total RC to muon capture
derived in Ref.~\cite{Czarnecki:2007th} and applied here are in
agreement, but its overall
uncertainty has been reduced by a factor of 4 in the present paper.

\subsection{Inputs \label{sec:inputs}} 

\begin{table}[htp]
\caption{Input parameter values used in this paper.  See text for
  discussion.
  }
\begin{center}
\begin{tabular}{ccccc}
  \hline\hline
 Symbol  	&Description	&Value & Source/Reference\vspace{1mm}\\ 
\hline
$G_F$ &  Fermi coupling constant	&	$1.1663787(6) \!\times\!10^{-5}\,{\rm GeV}^{-2}$ 	& Muon lifetime/PDG~\cite{Patrignani:2016xqp}\\
$V_{ud}$		&	CKM matrix element	&
                                                          0.97420(18)(10)		& Superallowed $\beta$ decays~\cite{Hardy:2014qxa}		\\
$f_\pi$		&	pion decay constant	&	92.3(1)\,{\rm MeV}	&	PDG~\cite{Patrignani:2016xqp}	\\
\gpinn		&	pion nucleon coupling 	&	13.12(10)	&	\cite{Baru:2010xn,Baru:2011bw}	\\
\hline
 \multicolumn{4}{l}{expansion parameters for nucleon charged current
  form factors}  \vspace{0.5mm}\\
$r_{1}^2$		&	squared rms radius for \FV	&
                                                                  0.578(2)
                                                                  fm$^2$
                                       & $n$-$e$, $\mu H$, see text
\\
\FM(0)		&	weak magnetic coupling	& \FMN	
                                       & see text	\\
$r_{2}^2$		&	squared rms radius for \FM	&	0.707(53) fm$^2$ &
$e$-$p$, $e$-$n$, $\pi$-$N$, see text
\\

$g_A \equiv F_A(0)$		&	axial coupling	&	\gaN	&
                                                                  $\tau_n$, see text
\\
$r_{A}^2$		&	 squared rms radius for \FA	&
                                                                  \raqN
                                                                  fm$^2$
                                       & $\nu - d$~\cite{Meyer:2016oeg}
  \\
\hline
 \multicolumn{4}{l}{derived nucleon charged current form factors at \qqC} \vspace{0.5mm} \\
\FV(\qqC)		&	vector form factor	&\FVCN	& this
                                                                  work\\
\FM(\qqC)		&	\hspace{-8mm}weak magnetic form factor\hspace{-8mm}	&\FMCN     & this
                                                                  work\\
$\bar{g}_A\equiv\FA(\qqC)$		&	axial form factor	& \gaCN& this work\\
	                                                          
$\!\!\!\bar{g}_P\equiv \frac{m_\mu}{m_N} \FP(\qqC)\!\!\!$ 		&	\hspace{-4mm}pseudoscalar form factor\hspace{-4mm}	& \gpCN& this
                                                                  work\\
\hline\hline
\end{tabular}
\end{center}
\label{tab:input}
\end{table}

The relevant inputs used to compute the capture rate are displayed in Table~\ref{tab:input}.
The Fermi constant $G_F$ is determined from the muon lifetime~\cite{Tishchenko:2012ie} and its
uncertainty is negligible in determining the muon capture rate. 
The CKM matrix element $|V_{ud}|$ is determined from superallowed $\beta$ decays~\cite{Hardy:2014qxa}.
The uncertainty in Table~\ref{tab:input} is divided into a nucleus-independent radiative correction
term, $0.00018$, and a second term $0.00010$ representing the sum in quadrature of other theoretical-nuclear
and experimental uncertainties.
The former radiative correction is strongly correlated with RC(electroweak) in Eq.~(\ref{eq:ew_radcor}),
and the corresponding uncertainty largely cancels when the muon capture rate is expressed in terms of
$|V_{ud}|$.  This cancellation has been accounted for in our discussion of
radiative corrections; in the numerical analysis the uncertainty contribution $0.00018$ to $|V_{ud}|$
is dropped.%
\footnote{The muon capture rate could be expressed directly in terms of $\beta$ decay observables,
  such as the neutron lifetime and superallowed beta
decay ft values, where $|V_{ud}|$  does not appear explicitly.}

The charged current isovector vector form factors are obtained from the isovector
combination of electromagnetic form factors.
Deviations from $F_1(0)=1$ occur at second order in small isospin
violating quantities.  At the quark level these quantities may be identified
with the quark mass difference $m_u - m_d$ and the electromagnetic coupling $\alpha$. 
At the hadron level, isospin violation manifests itself as mass splittings within multiplets, 
such as isodoublet $m_n-m_p$ and isotriplet $m_{\pi^\pm}^2-m_{\pi^0}^2$~\cite{Behrends:1960nf,Terentev1963,Ademollo:1964sr}.
As shown in Ref.\cite{Behrends:1960nf}, first-order isodoublet mass splitting
corrections vanish in $F_1(q^2)$ and $F_2(q^2)$, for general $q^2$,
while first order isotriplet ones
cancel in $F_1(0)$ but contribute in $F_1(q^2)$ for $q^2 \ne 0$ and in $F_2(q^2)$
for all values of $q^2$.
Estimating these corrections to be of $\order([m_{\pi^+}^2 - m_{\pi^0}^2]/
m_{\rho}^2) = 2.1 \times 10^{-3}$, where $m_\rho \approx 770\,{\rm MeV}$ is the $\rho$ meson mass (representing a typical
hadronic mass scale), 
we note that in $F_1(q_0^2)$ they are
accompanied by the further suppression factor $q_0^2 \;r_1^2/6 = -2.4 \times
10^{-2}$, so they amount to $- 5 \times 10^{-5}$. Corrections to the
isospin limit in $F_1(q_0^2)$ are thus negligible at the required permille level.
In the case of $F_2(q_0^2)$, we note that in the expression for the
singlet capture rate [Eq.~(\ref{eq:withFF}) below], a $2.1 \times 10^{-3}$ correction to
the $F_2$ term within square brackets 
amounts to $6.67 \times 10^{-4}$, while the total contribution from the 
four form factors is 4.217.%
\footnote{The additional suppression may be traced to a 
  factor $m_\mu/m_N$ appearing in the coefficients of $F_2$ relative to $F_1$ in Eq.~(\ref{eq:ci}).
  A similar power counting applies to the second class form factors, $F_S$ and $F_T$ in
  Eq.~(\ref{eq:VA}), that we have neglected in our analysis.}
Thus, a $2.1 \times 10^{-3}$ isotriplet
mass splitting correction to 
$F_2(q_0^2)$ induces a $2 \times 6.67 \times 10^{-4}/4.217 = 3.2 \times
10^{-4}$ correction to the singlet capture rate, which is also
negligible at the permille level.

Neglecting these small corrections, 
the Dirac form factor is thus normalized to $F_1(0) = 1$.  The Pauli form factor
at zero momentum transfer is given by the difference of the proton and neutron
anomalous magnetic moments: $F_2(0)=\kappa_p-\kappa_n$, where 
$\kappa_p= 1.79408$ and $\kappa_n = -1.91436$ are measured in units of  $e/2 m_N$. 
This leads to $F_2(0) = \FMN$. Note that since the PDG~\cite{Patrignani:2016xqp} expresses both
proton and neutron magnetic moments in units of $e/2m_p$, our value for $F_2(0)$ differs
from a simple difference of magnetic moments quoted there by a factor $m_N/m_p = 1.00069$.

The $q^2$ dependence of the form factors is encoded by the corresponding radii,
defined in terms of the form factor slopes:
\begin{linenomath*}
\begin{align}\label{eq:raddef}
{1\over F_i(0)}{dF_i \over dq^2}\bigg|_{q^2=0} \equiv \frac16 r_i^2 \,. 
\end{align}
\end{linenomath*}
Curvature and higher-order corrections to this linear approximation enter at second
order in small parameters $q_0^2/\Lambda^2 \sim m_\mu^2/m_\rho^2$,
where $\Lambda$ is a hadronic scale characterizing the form factor.  These
corrections may be safely neglected at the permille level.
Isospin violating effects in the determination of the radii may be similarly neglected. 
The Dirac-Pauli basis $F_1$, $F_2$ is related to the Sachs electric-magnetic basis $G_E$, $G_M$
by $G_E= F_1 + (q^2/4m_N^2)F_2$, $G_M=F_1+F_2$.  In terms of the corresponding
electric and magnetic radii,%
\footnote{The isovector form factors can be written in the form $F_i = F_{i,p} - F_{i,n} (i=1,2)$,
  $G_E = G_{E,p} - G_{E,n}$, $G_M = G_{M,p} - G_{M,n}$, where the subscripts $p$ and $n$
refer to the proton and neutron contributions. The electric and magnetic radii are defined analogously to Eq.~(\ref{eq:raddef})
in terms of the slopes
of $G_{E,p}$, $G_{E,n}$, $G_{M,p}$ and $G_{M,n}$. For the neutron, with $G_{E,n}(0) = 0$,
$r_{E,n}^2 \equiv 6\; G^\prime_{E,n}(0)$.}
\begin{align}
  r_1^2 = r_{E,p}^2 - r_{E,n}^2 -\frac{3}{2 m_N^2} F_2(0) \,,
  \quad 
r_2^2 = \frac{1}{F_2(0)} (\kappa_p \; r^2_{M,p} - \kappa_n\; r^2_{M,n} - r_1^2) \,. 
\end{align}
The neutron electric radius is determined from neutron-electron scattering length measurements,
$r_{E,n}^2 = -0.1161(22)\,{\rm fm}^2$~\cite{Patrignani:2016xqp}.
The proton electric radius is precisely determined from muonic hydrogen
spectroscopy, $r_{E,p} = 0.84087(39)\,{\rm fm}$~\cite{Antognini:1900ns}; this result remains controversial,
and is $5.6\sigma$ discrepant with the value $r_{E,p}=0.8751(61)\,{\rm fm}$ obtained
in the CODATA 2014 adjustment~\cite{Mohr:2015ccw}
of constants using electron scattering and ordinary hydrogen spectroscopy.
We take as default the more precise muonic hydrogen value, but verify that this $r_{E,p}$ puzzle
does not impact the capture rate at the projected 0.33\% level.
The magnetic radii are less well constrained.  We adopt the values
$r_{M,p} = 0.776(38)\,{\rm fm}$~\cite{Patrignani:2016xqp}
and $r_{M,n}=0.89(3)\,{\rm fm}$~\cite{Epstein:2014zua}.%
\footnote{This PDG value for $r_{M,p}$ represents the $z$ expansion reanalysis~\cite{Lee:2015jqa}
  of A1 collaboration electron-proton scattering
  data~\cite{Bernauer:2013tpr}.  A similar reanalysis of other world data
  in Ref.~\cite{Lee:2015jqa} obtained $r_{M,p}=0.914(35)\,{\rm fm}$.  We verify that this
  $r_{M,p}$ discrepancy does not impact the capture rate at the projected 0.33\% level.
  For $r_{M,n}$, we adopt the value from the $z$ expansion reanalysis~\cite{Epstein:2014zua} of
  $G_{M,n}$ extractions, combined with dispersive constraints (see also Ref.~\cite{Hill:2010yb}). 
  The larger uncertainty encompasses the PDG value, $0.864^{+0.009}_{-0.008}\,{\rm fm}$, obtained
  by averaging with the dispersion analysis of Ref.~\cite{Belushkin:2006qa}.}

 Currently, the most precise determination of $g_A$ comes indirectly
 via the neutron lifetime, $\tau_n$, used in conjunction with $V_{ud} = 0.97420(18)(10)$
 obtained from super-allowed nuclear $\beta$ decays~\cite{Hardy:2014qxa,Marciano:2005ec,Czarnecki:2004cw}.
 Correlating theoretical uncertainties in the electroweak radiative corrections
 to $\tau_n$ and $V_{ud}$,%
 \footnote{The first, $1.8\times 10^{-4}$, uncertainty on $|V_{ud}|$ in Table~\ref{tab:input}
   is correlated with the $1.9\,{\rm s}$ uncertainty on the right hand side of Eq.~(\ref{eq:gA_Vud_taun}). 
   These uncertainties cancel. 
 }
 reduces the uncertainty in Eq.~\eqref{eq:gA_Vud_taun} to 
\begin{equation}
1+3g_A^2 = 5172.0(1.1)\,{\rm s}/\tau_n \; 
\label{eq:taun}
\end{equation}%
In this review we use the average lifetime $\tau_n = 879.4(6)\,{\rm s}$~\cite{Czarnecki2018} from the UCN storage experiments (applying a
scale factor of S=1.5 according to the PDG convention~\cite{Patrignani:2016xqp}) and do not include the larger lifetime
measured in beam experiments. This choice is motivated by two additional and rather precise 
$\tau_n$ measurements with trapped neutrons~\cite{Serebrov2017,Pattie:2017vsj} since the last PDG evaluation, and the excellent agreement
of the average UCN storage lifetime with the more recent neutron decay
asymmetry measurements~\cite{Brown2017,Czarnecki2018}.  
Employing Eq.~\eqref{eq:taun} yields $g_A=\gaN$, which we use throughout this paper.
The current $g_A=1.2731(23)$ from  neutron decay asymmetry measurements, updated with the new
UCNA result~\cite{Brown2017}, is lower
and has nearly 5-times larger uncertainty, partially due to the large error
scaling factor S=2.3 caused by the inconsistency between earlier and post-2002
results.
On the other hand, if one includes only PERKEO and UCNA values, in particular
the recent preliminary PERKEO III result,%
\footnote{H.~Saul, International Workshop on Particle Physics at Neutron Sources 2018, ILL, 2018.}
the asymmetry measurements provide
a consistent $g_A$ determination of equal precision to the one used in this paper.

Our knowledge about the functional form of $F_A(q^2)$ relies primarily on
neutrino-deuteron scattering data from bubble chamber experiments in
the 1970's and 1980's: the ANL 12-foot deuterium bubble chamber
experiment~\cite{Mann:1973pr,Barish:1977qk,Miller:1982qi}, the BNL
7-foot deuterium bubble chamber experiment~\cite{Baker:1981su}, and
the FNAL 15-foot deuterium bubble chamber
experiment~\cite{Kitagaki:1983px,Kitagaki:1990vs}. 
As mentioned in the Introduction, 
the original analyses and most follow-up analyses employed the 
one-parameter dipole model of the axial form factor.  A more realistic
assessment of uncertainty allows for a more general functional
form.
Using a $z$ expansion analysis~\cite{Meyer:2016oeg}, the uncertainty
on the axial radius is found to be significantly larger than from
dipole fits,
\begin{linenomath*}
\begin{align}
  r_A^2(z\,\, {\rm exp.},\; \nu ) &= \raqN \, {\rm fm}^2 \,. 
\label{eq:raqz}
\end{align}
\end{linenomath*}
That analysis also investigated the dependence of the extracted axial radius on
theoretical statistical priors (different orders of truncation for the $z$ expansion,
the range over which expansion coefficients were allowed to vary, and choice of the
parameter $t_0$ defined by $z(t_0)=0$), finding that these variations are subdominant in the
error budget.  Radiative corrections were not incorporated in the
original experimental analyses.  
Radiative corrections and isospin violation would contribute percent
level modifications to the kinematic distributions from which \raq\
 is extracted, well below the statistical and systematic uncertainties of the existing datasets.
The value in Eq.~\eqref{eq:raqz} may be compared to a fit of scattering data to the
dipole form, $r_A^2({\rm dipole},\;\nu) =\raqdipnuN\,{\rm fm}^2$~\cite{Bodek:2007ym}.
Note that the value $r_A^2({\rm dipole})=0.454(13)\,{\rm fm}^2$ quoted in the
Introduction is obtained by averaging this neutrino scattering result
with an extraction from pion electroproduction~\cite{Bodek:2007ym},
$r_A^2({\rm dipole,\;electro.}) =0.454(14) \,{\rm fm}^2$.
As observed in Ref.~\cite{Bhattacharya:2011ah}, the electroproduction extraction
is also strongly influenced by the dipole assumption.  A more detailed discussion of
the electroproduction constraints is given in Sec.~\ref{sec:impact}, with the
conclusion that further control over systematics is required in order to provide a
reliable $r_A^2$ extraction. 

The pion decay constant $f_\pi$ and pion nucleon coupling $g_{\pi NN}$, along with $r_A^2$, are used to determine
the induced pseudoscalar form factor~\cite{Bernard:1994wn} 
\begin{linenomath*}
\begin{equation}
\label{eq:gp}
  \text{\FP}(q_0^2)
  = \frac{2\;m_N\; g_{\pi NN}\; f_\pi} {m_\pi^2-q_0^2}
-\frac{1}{3}\, g_A \, m_N^2\, r_A^2 + \dots \,,
\end{equation}
\end{linenomath*}
where $m_\pi = 139.571\,{\rm MeV}$ is the charged pion mass. 
Two loop \hb\ corrections, indicated by the ellipsis in Eq.~\eqref{eq:gp}, were
estimated to be negligible, as long as the low energy constants involved remain at natural
size~\cite{Kaiser:2003dr}.
$f_\pi$ is determined from the measured rate for $\pi^-\to \mu^- \bar{\nu}_\mu (\gamma)$,
and its uncertainty is dominated by hadronic structure dependent radiative corrections.
For $g_{\pi NN}$ we take as default the
value $g_{\pi NN} = 13.12(6)(7)(3) = 13.12(10)$~\cite{Baru:2010xn,Baru:2011bw},
where the first two errors are attributed to pion-nucleon scattering phase shifts
and integrated cross sections, respectively, entering
the Goldberger-Miyazawa-Oehme (GMO) sum rule for $g_{\pi NN}$. 
The third error is designed to account for isospin violation and was
motivated by evaluating a subset of \hb\ diagrams.
Other values include $g_{\pi NN} = 13.06(8)$~\cite{deSwart:1997ep},
%
%$g_{\pi NN} = 13.25(5)$~\cite{Perez:2016aol}
%
$g_{\pi NN} = 13.25(5)$~\cite{Perez:2016aol}
from partial wave analysis of
nucleon-nucleon scattering data; and
$g_{\pi NN} = 13.14(5)$~\cite{Arndt:2003if}, 
$g_{\pi NN} = 13.150(5)$~\cite{Arndt:2006bf}
from partial wave analysis of pion-nucleon scattering data.
That range of values is covered by the error given in Table~\ref{tab:input}.

\subsection{Numerical results}

Employing the radiative corrections given above, the full capture rates
become
\begin{linenomath*}
\begin{equation}\label{eq:LambdaRC}
  \Lambda = \left[ 1+ {\rm RC}  \right] \Lambda_{\rm tree} =  
  \left[ 1 + 0.0277(4)(2) - 0.005(1) \right] \Lambda_{\rm tree} \,,
\end{equation}
\end{linenomath*}
where $\Lambda_{\rm tree}$ is the tree level expression for the chosen spin state.
We have displayed a conventional separation of the radiative corrections in
Eq.~(\ref{eq:LambdaRC}), where 
the first~$+2.77\%$ includes the electroweak
 and electron vacuum polarization corrections, and the second $-0.5\%$ 
is the finite size correction. 
Inserting the relevant quantities from Table~\ref{tab:input},
the singlet 1S capture rate is given by
\begin{linenomath*}
\begin{equation}
\LS= 40.229 (41)\; [\FV(\qqC)+0.08833\; \FM(\qqC)+2.63645\; \gaC - 0.04544\; \gpC]^2 \; \si \,,
\label{eq:withFF}
\end{equation}
\end{linenomath*}
where the quantities $\gpC$ and $\gaC$ are defined below
and the relative uncertainty \ur\ = 1.0$\times 10^{-3}$ in the prefactor of Eq.~(\ref{eq:withFF})
quadratically sums the relative uncertainties \ur(RC) = (1.0, 0.4 and 0.2)$\times 10^{-3}$ 
and  (0.4 and 0.2)$\times 10^{-3}$ resulting from \ur($|V_{ud}|^2$) , taking into account that the two 0.4$\times 10^{-3}$
uncertainties are anticorrelated and cancel each other.

In the discussion above, we define $u_r= {\delta X}/{X}$ as the relative
uncertainty in the considered quantity  $X$ having an
uncertainty $\delta X$. The relative uncertainty in $X$ induced by 
parameter $p$ with uncertainty  $\delta p$ is $ u_r(p)={X}^{-1}
({\partial X }/{\partial p }) \delta p$. 

As a  next step, we evaluate the
form factors at the momentum transfer $\qqC$ relevant
for muon capture. For the vector form factors, we
expand to linear order using Eq.~(\ref{eq:raddef}), 
\begin{linenomath*}
\begin{align}
  \FV(\qqC) = \FVCN\,, \quad
  \FM(\qqC) = \FMCN  \,. 
\end{align} 
For the axial form factor we have 
\begin{equation}
\gaC \equiv \FA(\qqC) = 1.2510\; (118)_{\raq}\; (5)_{\ga} = \gaCN \,, 
\end{equation}
\end{linenomath*}
with the uncertainty dominated by \ur(\raq) = $9.4\times 10^{-3}$.
Finally, the pseudoscalar form factor predicted by \hb\ is
\begin{linenomath*}
\begin{equation}\label{eq:gpCdef}
\gpC \equiv {m_\mu \over m_N} \FP(\qqC) = 8.743\; (67)_{\gpinn}(9)_{f_\pi} -
0.498\; (238) _{\raq}\;
= \gpCN \; ,
\end{equation}
\end{linenomath*}
where the contribution from the pole and higher order term in
Eq.~\eqref{eq:gp} are shown separately. While the pole term dominates
the value for \gpC, the uncertainty is actually dominated by the
non-pole term, due to the rather dramatically increased uncertainty in \raq. 

We exhibit the sensitivity to the axial form factors by inserting the
relatively well known vector form factors in Eq.~(\ref{eq:withFF}) to
obtain
\begin{linenomath*}
\begin{equation}
\LS= 67.323(70)\; \big[ 1.00000(56) + 2.03801\; \gaC - 0.03513\; \gpC \big]^2 \; \si \,. 
\label{eq:withAFF}
\end{equation}
\end{linenomath*}
At the central values for $\gaC$ and $\gpC$,
the uncertainty in this equation from the remaining inputs is $\delta\LS = 1.03\; \si$,
corresponding to a relative error \ur=1.0$\times 10^{-3}$, which is still dominated by RC,
with a minor contribution from \ur(\FM)=0.3$\times 10^{-3}$.
At this point the traditional approach would be to insert \gaC\ and
\gpC\ in the equation above 
and to specify the uncertainties in \LS\ arising from these two axial
form factors. However, as both \gaC\ and \gpC\ depend on the axial
radius squared \raq, which is not well known, they cannot be treated
as independent input quantities. To avoid their correlation, we
express \LS\ in terms of the independent input parameters (\ga, \raq,
\gpinn):
\begin{linenomath*}
\begin{equation}
\LS = 67.323(70) \; [1.00000(56) - 0.02341(3)\; \gpinn + (2.03801 - 0.05556\; \raq)\;  \ga]^2
\; \si \;,
\label{eq:independent}
\end{equation}
\end{linenomath*}
with $r_A^2$ in units of fm$^2$.
Using the current knowledge of these independent input quantities from
Table~\ref{tab:input}, we obtain our best prediction for the muon
capture rate in the singlet and triplet  hyperfine states of muonic
hydrogen as
\begin{linenomath*}
\begin{align}
  \LS &= 715.4 \; (6.9) \; \si  \,,
  \label{eq:final1}
\\
\LT &=12.10 \; (52) \; \si \;.
\label{eq:finalLT}
\end{align}
\end{linenomath*}
We have employed the same methodology as above for \LS\ to obtain \LT.   
The total relative uncertainty for $\LS$, \ur = 9.7$\times 10^{-3}$, is
calculated as the quadratic sum of \ur(RC) = 1.0$\times 10^{-3}$,
\ur($g_{\pi NN}$) = 1.4$\times 10^{-3}$, 
\ur(\ga) = 0.6$\times 10^{-3}$, \ur(\raq) = 9.5$\times 10^{-3}$ and a
negligible uncertainty from $f_\pi$.
Assuming no uncertainty in \raq,  the prediction for
\LS\ would have a more than 5 times smaller error of  $1.3\; \si$.

\section{Muon capture experiment update}
\label{sec:experiment}

\begin{figure}[ht]
  \begin{center}
    \includegraphics[scale=.5]{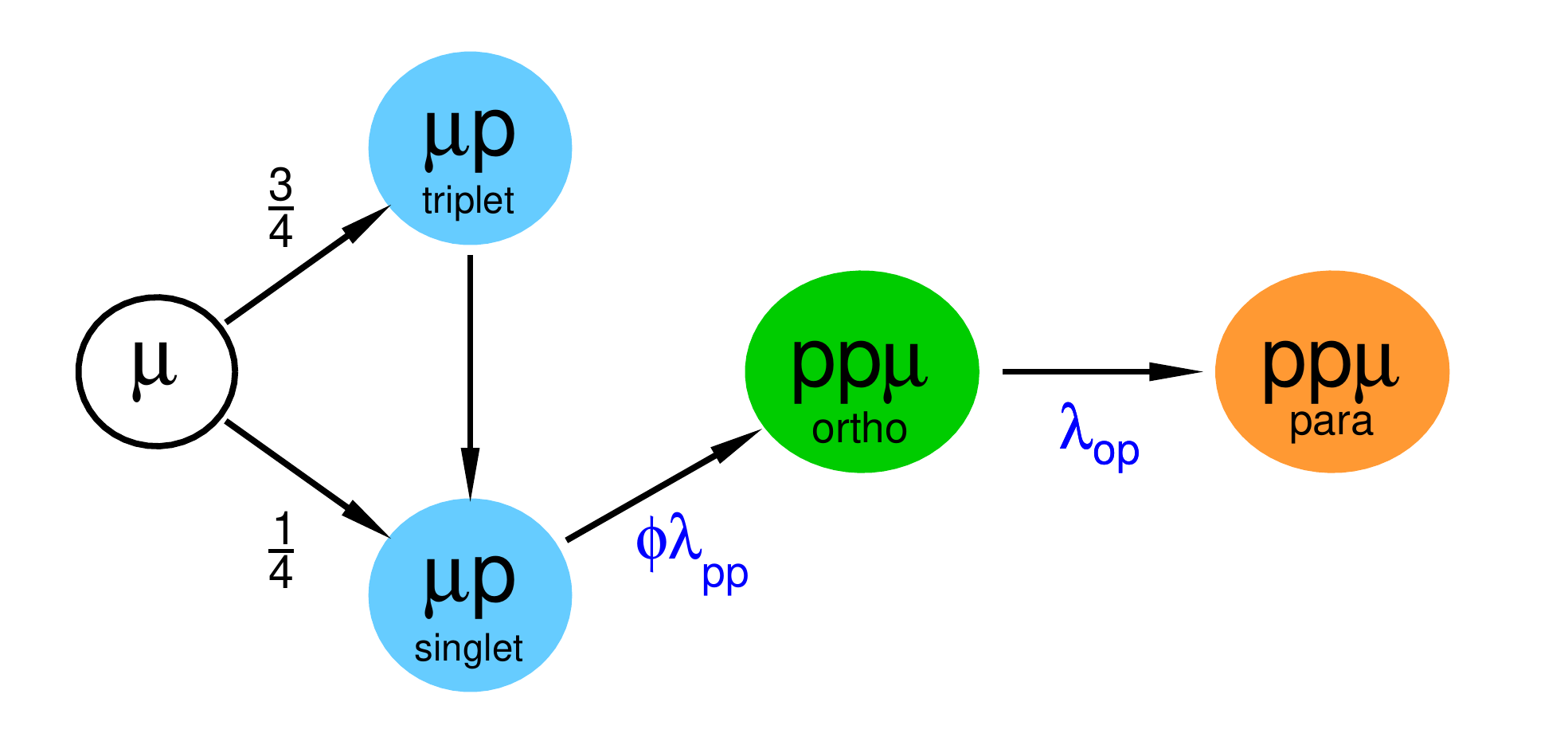}
    \caption{\label{fig:mup_kinetics} (color online). Reaction sequence after muons
              stop in hydrogen. Triplet states are quickly
              quenched to the singlet $\mu H$ ground
              state. In collisions, ($pp \mu$) ortho-molecules are formed
    proportional to the hydrogen density $\phi$ and the formation rate
$\lambda_{pp}$. Ortho-molecules can convert to para-molecules
with the poorly-known rate $\lambda_{op}$. Reproduced from \cite{Kammel:2010zz}.}
  \end{center}
\end{figure}

Precise measurements of muon capture in hydrogen are
challenging, for the following reasons~\cite{Kammel:2010zz,Gorringe:2002xx}.
(i) Nuclear capture takes place after muons come to rest in matter
and have cascaded down to the ground state of muonic atoms. As
exemplified for the case of $\mu$H in Eq.~\eqref{eq:rate}, the capture
rate is proportional to the square of the muonic wavefunction at the
origin,%
\footnote{Compare Ref.~\cite{Czarnecki1998} for corrections
  relevant for capture and muon-electron conversion in heavy nuclei.}
 $|\psi_{\rm 1S}(0)|^2$, which, after summing over the number
of protons in a nucleus of charge $Z$, leads to a steep increase of the
capture rate with $\sim Z^4$, such that the muon capture and decay
rates are comparable for $Z\sim$13. For $\mu$H, where $Z=1$, this
amounts to a
small capture rate of  order 10$^{-3}$ compared to muon decay, as
well as dangerous background from muon stops in other higher $Z$ materials, where
the capture rate far exceeds the one in $\mu$H.
(ii) On the normal atomic
scale $\mu$H atoms are small and can easily penetrate the
electronic cloud to  transfer
to impurities in the hydrogen target gas, or to form muonic molecular
ions $(pp\mu)^+$. 
The former issue requires
target purities at the part-per-billion level.
The latter problem, depicted
in Fig.~\ref{fig:mup_kinetics}, has been a primary source of confusion
in the past, as the helicity dependence of weak interactions implies
large differences in the capture rates from the possible states. The rates
for the two atomic hyperfine $\mu$H states are given in Eqs.~(\ref{eq:final1},\ref{eq:finalLT}),
while the molecular rates can be calculated as 
\begin{linenomath*}
\begin{align}
\Lambda_{\rm ortho} &= 544 \; \si \;, \\
\Lambda_{\rm para} &= 215 \; \si \;,
\label{eq:molecule}
\end{align}
\end{linenomath*}
using the molecular overlap factors given in Eq.~(11) of
Ref.~\cite{Kammel:2010zz}.%
\footnote{We do not estimate uncertainties for the
          molecular rates, as a reliable error evaluation at the permille
  level should include a modern confirmation of the original
  calculation of the $(pp\mu)^+$ space and spin structure~\cite{Bakalov:1980fm}.}
To interpret a specific experimental capture
rate, the fractional population of states for the given experimental
conditions has to be precisely known, which is especially problematic
for high density targets. 
(iii) Finally, muon capture in hydrogen leads to an all neutral final state, $n+\nu$,
where the 5.2 MeV neutron is hard to detect with well-determined
efficiency.

\subsection{MuCap experiment: strategy and results}

Over the past two decades, the $\mu^3$He experiment, the MuCap and
later the MuSun collaboration have
developed a novel active target method based on high pressure time
projection chambers (TPC) filled with pure $^3$He, ultra-pure hydrogen (1\% of
liquid hydrogen (LH$_2$) density) or cryogenic deuterium gas (6\% of LH$_2$ density),
respectively, to overcome the above challenges. The first experiment~\cite{Ackerbauer:1997rs},
benefiting from the charged final state, 
determined the rate for $\mu+ {}^3{\rm He}\rightarrow t + \nu$ with an
unprecedented precision of 0.3\% as $1496.0\pm 4.0\, \si$. The most
recent extraction~\cite{Marcucci:2011jm} of \gpC\ from this result
gives $\gpC=8.2(7)$, with uncertainties
due to nuclear structure theory. Additional uncertainties would
enter if the new $r_A^2(z\; {\rm exp.},\; \nu)$ is taken into account.%
\footnote{We refrain from updating this
          result with new form factors. As the calculation uses
          tritium decay as input, changes in form factors at \qq=0 are
          expected to cancel, but the uncertainty in the momentum
          dependence enters via \raq.}

MuCap measured  \LS\  in the theoretically clean $\mu$H system to
extract \gpC\ more directly.
The original publication~\cite{Andreev:2012fj} gave 
$\Lambda_{\rm singlet}^{\rm MuCap}=(714.9 \pm 5.4_{\rm stat} \pm
5.1_{\rm syst}) \text{ s}^{-1}$,
which was slightly updated based on an improved determination of  the
($pp \mu $) molecular formation rate $\lambda_{pp}$~\cite{Andreev:2015evt}
to its final value
\begin{equation}\label{eq:exp_update}
\Lambda_{\rm singlet}^{\rm MuCap}=(715.6 \pm 5.4_{\rm stat} \pm 5.1_{\rm syst}) \text{ s}^{-1} \,. 
\end{equation}
The scientific goal of MuSun~\cite{Andreev:2010wd,Kammel:2010zz} is the determination of an important low
energy constant (LEC), which characterizes the strength of the axial-vector
coupling to the two-nucleon system and enters the calculation of
fundamental neutrino astrophysics reactions, like $pp$ fusion in the sun
and $\nu d$ scattering in the Sudbury Neutrino Observatory~\cite{Chen:2002pv}. 

As muon capture involves a characteristic momentum transfer of the order
of the muon mass, extractions of form 
factors and LECs from all of these experiments are sensitive to the modified theoretical
capture rate predictions or uncertainties implied by the use of the
new
$r_A^2(z\; {\rm exp.},\; \nu)$.

\begin{figure}[h!]
  \begin{center}
    \includegraphics[scale=.5]{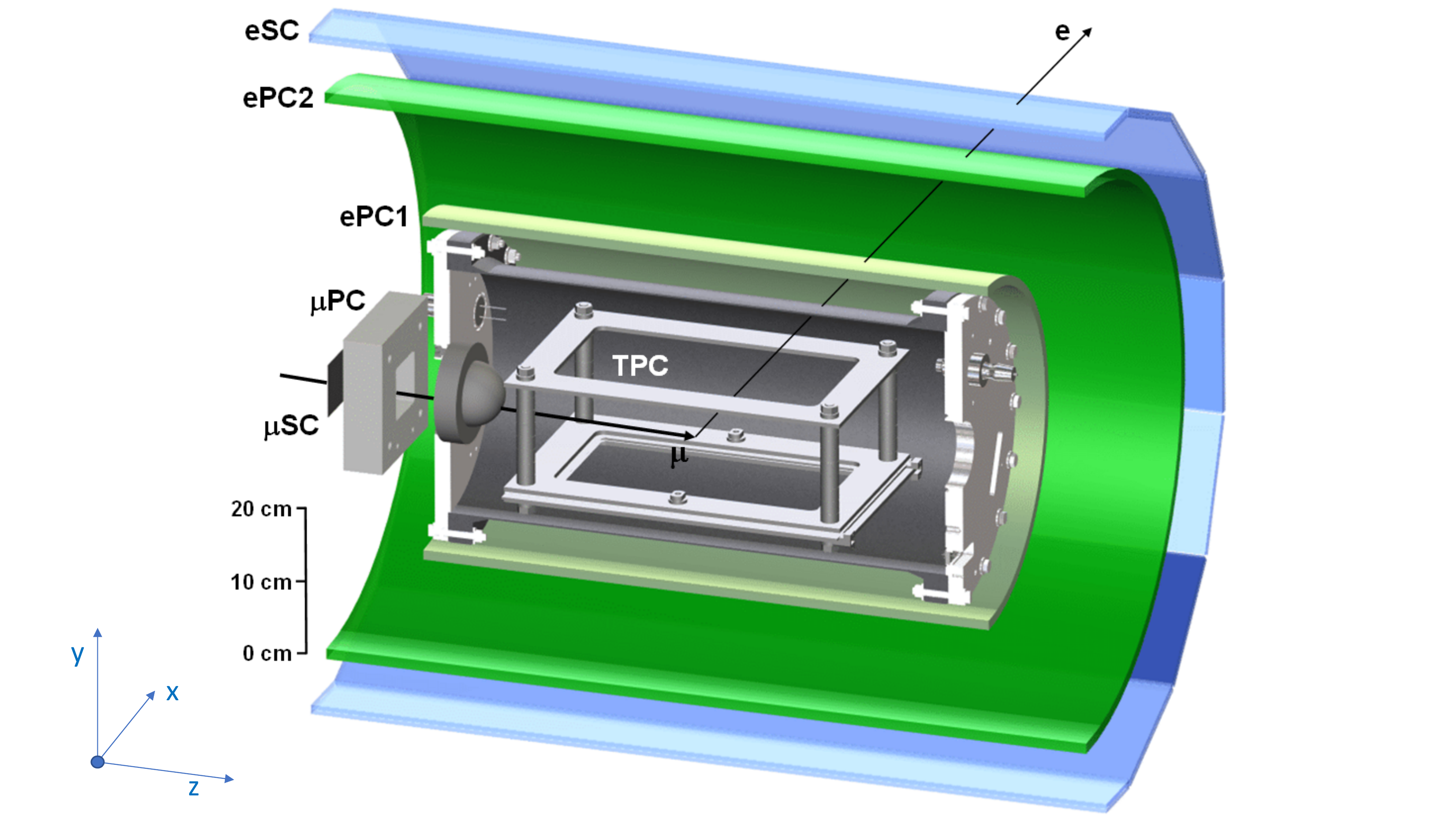}
    \caption{\label{fig:mucap} (color online) Simplified MuCap detector
              model. Reproduced from \cite{Andreev:2007wg}.  }
  \end{center}
\end{figure}

In view of potential further improvements, let us analyze in some
detail how MuCap achieved its high precision  1\% measurement. 
Fig.~\ref{fig:mucap} illustrates the basic concept. Muons
 are detected by entrance detectors, a 500-$\mu m$ thick scintillator
 ($\mu$SC) and  a wire chamber ($\mu$PC), and pass through a 
500-$\mu m$-thick hemispherical beryllium pressure window to stop in
the TPC,  
which is filled with ultrapure, deuterium-depleted hydrogen gas at a
pressure of 1.00 MPa and at ambient room temperature.
Electrons from muon decay are tracked in two cylindrical wire chambers 
($e$PC, green) and a 16-fold segmented scintillator array ($e$SC, blue). The
experimental strategy involves the following key features.

Low density and suppressed \ppm\ formation:
As the target has only 1\% of
liquid hydrogen density, molecule
formation is suppressed and 97\% of muon capture occurs in the $\mu$H
singlet atom, providing unambiguous interpretation of the signal.

Lifetime method~\cite{Bardin:1980mi}: The observable  is the disappearance rate $\lambda_- $ of negative
muons in hydrogen, given by the time between muon entrance and decay
electron signal. The capture rate is extracted as the difference $\LS
\approx \lambda_- - \lambda_+ $, where $\lambda_+$  is
 the precisely known positive muon decay
 rate~\cite{Tishchenko:2012ie}. Contrary to the traditional method of detecting capture neutrons from
process~\eqref{eq:omc} which requires absolute efficiencies,
only precise time measurements are needed, albeit at large statistics.

Selection of muon stops in hydrogen by tracking:
The TPC~\cite{Egger:2014bua} tracks the incident muons in
three dimensions to accept only hydrogen stops sufficiently far away from
wall materials with higher capture rate. Its sensitive volume is
$15 \times 12 \times 28 \;{\rm cm}^3$, with an electron drift velocity of 5.5
mm/$\mu$s at a field of 2 kV/cm in vertical $y$-direction. The 
proportional region at the bottom of the chamber was operated at a gas
gain of 125 -- 320, with anode (in $x$-direction)
and cathode (in $z$-direction) wires read out by time-to-digital
converters using three different discriminator levels.

Ultra-pure target gas: Target purity of $\sim10$~ppb was maintained
with a continuous circulation and filter system~\cite{Ganzha:2007uk}. The TPC allowed
in-situ monitoring of impurities by observing charged nuclear recoils
from $\mu^- + {\rm O} \rightarrow \nu_\mu + {\rm N}^*$, in the rare
cases  
of muon transfer to impurities, occuring at the 10$^{-5}$ level. Isotopically pure protium was produced
onsite~\cite{Alekseev:2015} and verified by accelerator mass spectroscopy~\cite{2008NIMPB.266.1820D}.
In total, $1.2\times 10^{10}$ decay events were accepted
with muons stopping in the selected restricted fiducial volume.

\subsection{Conceptual ideas towards a 3-fold improved muon capture experiment}

The 3-fold uncertainty reduction over MuCap implies a precision
goal of $\delta \LS \sim 2.4\, \si$ which, for definiteness, we assume equally
shared between $\delta \LS({\rm stat})=\delta \LS({\rm syst})=1.7~\si$. Achieving
this goal is no small feat, and we hope that the motivation for a low-$q^2$ measurement of the
axial form factor outlined in this paper will stimulate further
innovative experimental ideas. However, in the remainder of this section we
follow the more conservative approach to consider incremental
improvements to the MuCap strategy only. We note that MuCap was a
pioneering experiment developing a new technology, so it is likely
that a next generation experiment can be further optimized, based on
the lessons learned.

\subsubsection{Statistics}

A reduction of the 5.4~\si\ MuCap statistical uncertainty to 1.7~\si\
requires about a 10-fold increase in statistics. Typically, such
order of magnitude advances in nuclear/particle physics experiments require
concerted upgrades in beam and detector performance. 

MuCap accepted only events with a single muon entering the TPC
separated from neighboring muons by at least $T_{\rm obs}=25~\mu {\rm s}$. This
eliminated combinatorial distortions to the measured time spectra.
In the PSI continuous beam with a rate $R_{\mu}$, the rate for
those single muon events would be $R_{\mu} e^{-R_{\mu} 2\, T_{\rm obs}}$
only,  which is limited to roughly 7 kHz.  Thus a muon-on-request scheme was
developed (c.f. Fig.~\ref{fig:mucap}): A muon in the beam scintillator
$\mu$SC triggered a fast kicker~\cite{Barnes:2004}, which deflects the beam for the
measuring period $T_{\rm obs}$ to avoid muon pile-up. With $R_{\mu}= 65~\mbox{kHz}$, 
$\mu$SC had a pile-up free rate of 22~\mbox{kHz}, of which a
fraction $\epsilon_{\rm fid} \approx 0.3$
stopped in the fiducial volume of the TPC selected for physics
analysis. Including the electron detection efficiency of
$\epsilon_e=0.5$ and deadtime losses, the rate for accepted events was
$R_{\rm acc}\sim 2~\mbox{kHz}$.

To increase this rate, it is certainly worthwhile to explore whether the experiment could run
with multiple muons in the TPC. If this idea leads to
unacceptable systematic complications,  the single muon concept could
still be preserved by  increasing the muon stopping efficiency from
$\epsilon_{\rm fid}=0.3 \rightarrow 0.9$ and the electron detection
efficiency $\epsilon_e=0.5 \rightarrow 0.7$, together with supplemental
improvements in data taking efficiency. The resulting rate
increase of about 4.5 yields $R_{\rm acc}\sim 9~\mbox{kHz}$, 
so that $12 \times10^{10}$ events can be collected in 6 months of data
taking (including typical up-time fractions for beam and experiment).

We consider 3 main upgrades to reach this goal. i) Minimize any material
traversed by the muon beam, so that the beam momentum can be decreased
from 34 MeV/c to 29 MeV/c, which reduces the muon range by nearly a
factor 2.%
\footnote{This strong impact of the muon momentum p on its
          range R follows from the approximate relation $R \propto p^{3.6}$, which can be understood from integrating
          the Bethe-Bloch energy loss equation, c.f.  III.20 of \cite{Hikasa:1992je}.}
Since longitudinal range straggling, as well as part of the transverse expansion of the
beam, scales with the total range, a much more compact stopping
volume can be realized. The evacuated muon beam  pipe
should be directly connected to the TPC entrance flange, with the beam
detectors reduced to a 200 $\mu m$-thin $\mu$SC operating in vacuum
with modern silicon photomultiplier (SiPM) technology and the Be
window diameter reduced, so that it can be made thinner. The beam to air
windows and the wire chamber are
eliminated, the latter replaced by a retractable beam spot monitor,
which is only used during beam tuning and for systematic studies, but
does not add material during production data taking.
The beam pipe should be designed as a safety containment volume in case of
a breach of the Be window, by placing an additional thin window or
fast interlock in an upstream focus. In the first focus downstream of
the kicker, another thin scintillator might serve as the beam
trigger to minimize kicker delay. ii) With a collimated beam impinging
on $\mu$SC and the detector itself
positioned as close as possible to the TPC, a stopping
efficiency $\epsilon_{\rm fid}$ approaching
unity can be expected for muons seen by this detector. iii) As the beam rate for negative muons drops steeply with
momentum, more powerful PSI beams, existing or under development, should
be considered.

\subsubsection{Systematics}

An uncertainty goal of  $\delta \LS(\rm syst)=1.7~\si$ implies that the negative muon decay rate
$\lambda_-$ in hydrogen has to be measured at least at a precision of
3.7 ppm relative to muon decay. This poses unprecedented requirements
on the TPC track reconstruction, as no early to late effects over the
measuring interval $T_{\rm obs}$ distorting the decay spectrum are
allowed at this level.
The main systematic corrections enumerated in Table II
of Ref.~\cite{Andreev:2012fj} can be grouped into 4 distinct classes:

i) Boundary and interference effects: By definition, for an infinite TPC no
boundary effects (like wall stops, scattering and diffusion)  would
occur. Interference effects, on the other hand,
are generated by decay electrons, affecting the muon stop
reconstruction in a time dependent manner.
The experiment has to balance these two competing
systematic effects, by carefully selecting muons within a clean fiducial
volume, without introducing interference distortions.  For the final,
best MuCap run R07 their total uncertainty added up to $\delta \LS
= 3.3~\si$.
The obvious remedy for boundary effects is a larger TPC volume,
coupled with potential geometry improvements, as well as the reduction
in the beam stopping volume. While the dimensions perpendicular to the
TPC drift field are only constrained by practical considerations, the
drift time cannot be much longer than $T_{\rm obs}$, in order to avoid reducing the
acceptable beam rate. We expect that the TPC can be operated with 
drift fields up to 10 kV/cm, as demonstrated at higher density in
MuSun~\cite{Ryan:2014jwa}, which would double the drift velocity and allow drift distances
of 20 cm. To improve the tracking quality, a geometry of independent pads like
MuSun has proven advantageous, as the MuCap cathode wires strung in the
direction of the muon tracks provide very limited information.
As MuSun demonstrated, full  digitization of all signals instead of
simple threshold timing information, adds powerful tracking capabilities.

ii) Gas impurities:
For the R07 run,  transfer to gas impurities amounted to $\delta \LS
\sim 1~\si$. Gold coating of inner vessel surfaces, improved gas chromatography
(already achieved in MuSun) and/or spectroscopy and, most importantly, full digital readout of the
TPC signals~\cite{Ryan:2014jwa}, for in-situ detection of capture
recoils, should reduce
this uncertainty to below 0.1 \si.

iii) Electron detector effects: A significant uncertainty of $\delta \LS
\sim 1.8~ \si$ was included in the MuCap error budget, because
of incompletely understood
discrepancies between alternative electron track
definitions. Because diffusion processes in hydrogen introduce
systematic problems when applying tight vertex cuts, MuCap
concluded that precision tracking is not essential. Thus a new
experiment should use scintillators or scintillating fibers with
SiPM readout, which are simple, robust and more stable than
the wire chambers used in MuCap. Then an
instrumental uncertainty below 1 ppm similar to
MuLan~\cite{Tishchenko:2012ie} can be expected.

iv) \ppm\ molecular effects: Although capture from \ppm\
molecules amounts only to 3\% in 1 MPa hydrogen gas, the uncertainty
introduced by the inconsistent determinations of the ortho-para rate $\lambda_{\rm op}$~\cite{Kammel:2010zz} 
shown in Fig.~\ref{fig:mup_kinetics}, introduces a $\delta \LS \sim
1.8~ \si$ uncertainty ~\cite{Andreev:2015evt}. As shown in Fig.~2 of Ref.~\cite{Andreev:2012fj}, the
poor knowledge of $\lambda_{\rm op}$
also leaves unresolved the
question whether the previous measurement of ordinary muon
capture in liquid
hydrogen~\cite{Bardin:1980mi} or, alternatively, the measurement of
radiative muon 
capture~\cite{Wright:1998gi} strikingly deviates from theory. The high density cryogenic
TPC developed for the MuSun $\mu$D experiment, could settle both
issues with a first precise measurement of $\lambda_{\rm op}$ when filled
with protium gas of about 10\% liquid density.

Finally, it should be mentioned that the MuCap TPC occasionally
suffered sparking issues, which required running with reduced voltage.
Better stability and higher gain should be achieved by starting some
R\&D efforts with smaller prototypes, with improvements to the classical proportional
wire chamber technique used by MuCap as well as tests of --- now mature
--- micro-pattern
chamber alternatives, like GEMS and micro-megas.

\section{Results and opportunities}
\label{sec:reach}

Having reviewed the status of theory and explored the reach for experiment, 
in this section we evaluate how well the nucleon form factors and
coupling constants can be determined by the present MuCap experiment
at 1\% precision, and by a potential new experiment at the 0.33\% level.

\subsection{Updated value for the pseudoscalar coupling \gpC\ and
          extraction of \gpinn\ }

We begin our applications by using the final MuCap experimental result, $\Lambda_{\rm singlet}^{\rm MuCap} =715.6(7.4)\; {\rm s}^{-1}$,
together with our updated  $\Lambda_{\rm singlet}^{\rm theory}$  in Eq.~(\ref{eq:withAFF}), to extract a value for \gpC\ that can be compared with the prediction of \hb.
Both the experimental value and theoretical prediction depend on $r_A^2$.   To illustrate that dependence, we start with the
traditional value of $r_A^2({\rm dipole},\; \nu) = 0.453(23)\;{\rm fm}^2$
obtained from dipole fits to neutrino scattering data with a very small ($\sim 5\%$) uncertainty.
It leads to:

\begin{linenomath*}
\begin{align}
&\gpC^{\rm MuCap}\big|_{r_A^2=\raqdipnuN\,{\rm fm}^2} = 8.26\;  (48)_{\rm exp}\;(8)_{\gaC}\; (4)_{\rm RC}=  8.26 (49) \,,   &\gpC^{\rm theory}= 8.25(7) \;. 
\label{eq:gp_dipole}
\end{align}
\end{linenomath*}
For comparison, we take the ratio and find $\gpC^{\rm theory}/\gpC^{\rm MuCap} =1.00(6)$,
which exhibits very good agreement at the $\pm 6\%$ level.  
Alternatively, employing the more conservative $z$ expansion value obtained from neutrino scattering,
$r_A^2(z\; {\rm exp.},\; \nu)=0.46(22)\; {\rm fm}^2$,
with its nearly $50\%$ uncertainty, one finds:
\begin{linenomath*}
\begin{align}
&\gpC^{\rm MuCap}\big|_{r_A^2=\raqN\,{\rm fm}^2} = 8.23\; (48)_{\rm exp}\;(68)_{\gaC}\; (4)_{\rm RC}=
  \gpMuCapN \,,
  &\gpC^{\rm theory}= \gpCN \; , 
\label{eq:gp_z}
\end{align} 
\end{linenomath*}
where $\bar{g}_P^{\rm theory}$ is obtained from Eq.~(\ref{eq:gpCdef}). 
The uncertainties are considerably larger. However, taking the ratio and accounting for correlated errors,
$\gpC^{\rm theory}/\gpC^{\rm MuCap} =1.00(8)$. 
Agreement is still very good and theory is tested at about $\pm 8\%$,
not a significant loss of sensitivity. If $r_A^2$ could be independently determined with high precision
(for example, using lattice gauge theory techniques), then a new MuCap experiment
with a factor of 3 improvement would test \hb\ at about the 2\% level.

Alternatively, the measured capture rate in conjunction with the
theoretical formalism can be used to determine the  
pion-nucleon coupling \gpinn\ from the $\mu H$ atom. 
This approach is closely related to the extraction of the pseudoscalar
form factor, as \gpinn\ appears as the least well known parameter in
the PCAC pole term of Eq.~\eqref{eq:gp}. For this purpose
Eq.~\eqref{eq:withAFF} was recast in terms of the independent
parameters (\gpinn,  $g_A$ and $r_A^2$) into
Eq.~\eqref{eq:independent}, avoiding the correlation between the axial
form factors introduced by \raq. 
That prescription gives, for $r_A^2=0.46(22)\;{\rm fm}^2$:
\begin{linenomath*}
\begin{align}\label{eq:gpinn}
&g_{\pi NN}^{\rm MuCap} = \rm 13.11\; (72)_{exp}\;(4)_{\ga}\;
  (67)_{\raq}\;  (7)_{RC}=
               13.11 (99) \,,   &g_{\pi NN}^{\rm external}= 13.12(10) \;.
\end{align}
\end{linenomath*}
The result is in very good agreement with the external \gpinn\
obtained from pion-nucleon phase shift and scattering cross section data, such as the value given in Table~\ref{tab:input}.
It provides a direct 8\% test of \hb\, essentially the same as indirectly obtained from the
\gpC\ analysis given above.  As in the case of \gpC, a future factor of 3 improvement in the capture rate
combined with an independent precise determination of $r_A^2$ would
determine \gpinn\ to 2\%.

\subsection{Determination of \raq\ from muon capture \label{sec:rA2mu}}

The basic premise of this paper has been that the error on $r_A^2$ extracted from
neutrino scattering data is much larger (by about an order of magnitude)
than generally assumed. Indeed, the value~\cite{Meyer:2016oeg}
$r_A^2(z\;{\rm exp.}\,\nu) =0.46(22)\;{\rm fm}^2$,
based on the $z$ expansion method, that we employed, has a nearly 50\% uncertainty.
As we shall see in Sec.~\ref{sec:impact}, this is problematic for predicting quasi-elastic neutrino scattering
cross sections needed for next-generation neutrino oscillation studies. For that reason, it is
timely and useful to consider alternative ways of determining $r_A^2$. Various
possibilities are discussed in Sec.~\ref{sec:impact}; however, first we consider
existing and possible future implications from the MuCap experiment.

Muon capture provides a unique opportunity to determine \raq, highly
complementary to neutrino charged-current scattering. The momentum
transfer \qqC\ is small and well defined, rendering higher terms in
the \qqC\ Taylor expansion negligible. However, the effect of \raq\ is small,
with $F_A(\qqC)$ being only $\raq \,\qqC/6 \approx 2 \%$ smaller than
$F_A(0)$. Thus precision experiments at the sub-percent level are called for.

The change in \LS\ due to  a change in \raq\ is given in
Eqs.~\eqref{eq:withAFF},\eqref{eq:independent}, and can be quantified as
\begin{align}
\frac{\partial \LS}{\partial \raq} &=
                                                  \frac{\partial \LS
                                                  }{\partial \gaC }
                                \frac{\partial \gaC}{\partial \raq} +
                                                  \frac{\partial
                                                  \LS}{\partial \gpC}
                                                  \frac{\partial\gpC
                                                  }{\partial \raq} =
-47.8 + 16.7= -31.1 \;\rm \si\; fm^{-2}\;.
\end{align}
Thus, a one sigma step of $\rm 0.22\; fm^2$ in \raq\ changes \LS\ by
6.8 \si\ or about 1\%. Unfortunately, for the present purpose,  the sensitivity to the
axial radius is reduced, as the contributions from \gaC\ and \gpC\
counteract.

Employing Eq.\eqref{eq:independent} with
the input from Table~\ref{tab:input} we find
\begin{linenomath*}
\begin{equation}
\raq(\rm MuCap) = 0.46\; (24)_{exp}\;(2)_{\ga}\; (3)_{\gpinn}\;  (3)_{RC}=
               0.46 (24)\; \fmq. 
\end{equation}
\end{linenomath*}
This result is comparable in uncertainty to the $z$ expansion fit to the
pioneering neutrino scattering experiments~\cite{Meyer:2016oeg}. To
compare the two processes at higher precision would require inclusion
of  electroweak radiative corrections for the neutrino data at a level comparable to what has been done for muon capture.
Making the reasonable assumption that the two approaches are uncorrelated, we
can compute the weighted average
\begin{linenomath*}
\begin{equation} \label{eq:average}
\raq(\rm ave.) =  \raqAN\; \fmq.
\end{equation}
\end{linenomath*}
The averaged uncertainty has been reduced to about 35\%.
A future experiment, assumed to reduce the overall MuCap error from 1\% to 0.33\% would reduce the error in $r_A^2$ to
\begin{linenomath*}
\begin{equation}
\delta\raq(\rm \mbox{future exp.}) =  (0.08)_{exp}\;(0.02)_{\ga}\; (0.03)_{\gpinn}\;
(0.03)_{RC}= 0.09\; \fmq.
\end{equation} 
\end{linenomath*}
The muon capture squared axial radius determination, when averaged with the neutrino scattering  $z$ expansion result,
would then have about a 20\% uncertainty.  This precision level is
important, as it would be sufficient to reduce the \raq\ dependent
theoretical uncertainty in neutrino quasielastic cross sections 
to a subdominant contribution, as we demonstrate below in Sec.~\ref{sec:impact_rA2}.

If we would have used the updated average from the asymmetry measurements, $g_A=1.2731(23)$, the extracted \raq\ from MuCap
would be lower, $\raq (\rm MuCap)=0.38(25) \fmq$, but still easily within the current large error\footnote{The
  sensitivity of $r_A^2$ to $g_A$ is given by $\raq ({\rm MuCap}) \approx 0.46(24)+ 28.4( g_A -1.2756) \, {\rm fm}^2 $.}. 
For a 3-fold improved MuCap experiment,
the uncertainty introduced from that $g_A$ would be more serious, about the same as the experimental contribution to \raq. However, 
we expect that in the near future $g_A$ values will have converged,
which we antipate by using the more precise  $g_A=\gaN$ in this review.

\subsection{Determination of \ga\  and electron-muon universality \label{sec:universality}}

\begin{figure}[h!]
  \begin{center}
    \includegraphics[width=0.7\textwidth]{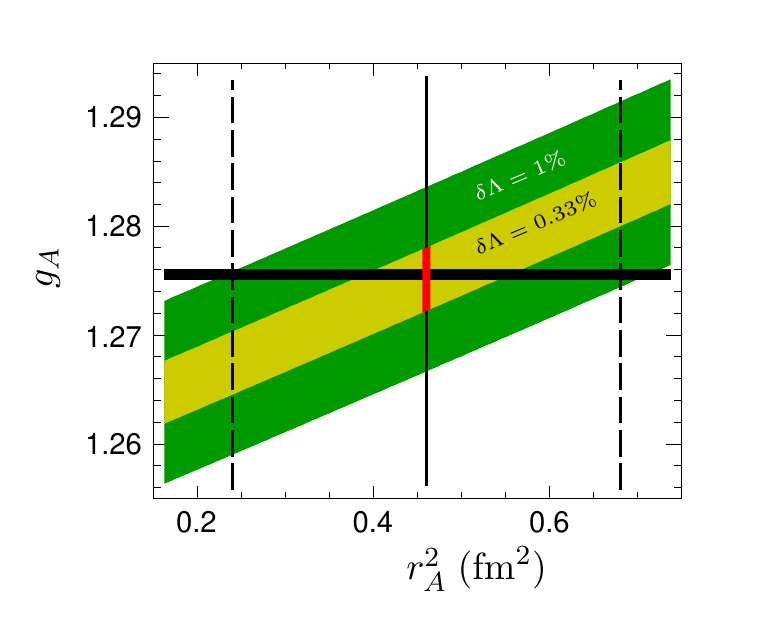}
    \caption{\label{fig:gA}
       (color online) Relation between \ga\ and \raq\  from electron and muon
      processes.  The horizontal black band shows  \ga\ from
      neutron $\beta$ decay (Table~\ref{tab:input}). The outer sloped green 
      band denotes the  $g_A-r_A^2$ region consistent with the present
      MuCap result within 1-sigma, the inner sloped yellow band the potential
     of a future 3-times improved measurement (the same central value
     has been assumed). The current value and uncertainty in \raq\ from the
     neutrino scattering analysis is shown by vertical lines. If \raq\
     would be known to 1\%, the future experiment would determine $g_A$
     within the vertical red bar.
    }
  \end{center}
\end{figure}
The axial coupling governing neutron $\beta$ decay, $g_A = F_A(0)$, 
is a critically important QCD induced physics parameter~\cite{Abele:2008zz}.  Taken together with the
neutron lifetime, $\tau_n$, it can provide a clean determination of $V_{ud}$
free of nuclear physics uncertainties, via Eq.~(\ref{eq:gA_Vud_taun}).
In addition, $g_A$ is needed for constraining the number of effective neutrino species from
primordial nucleosynthesis; computing reactor and solar neutrino fluxes and cross-sections;
parametrizing the proton spin content and testing the Goldberger-Treiman relation~\cite{Dubbers:2011ns}.
In this paper we use the value $g_A = \gaN$, 
derived from the chosen average $\tau_n $ of neutron bottle experiments and $V_{ud}$ given in Table~\ref{tab:input}
according to Eq.\eqref{eq:taun}.
The error on $g_A$ is expected to be further reduced to about $\pm 0.01 \%$,
by future $\tau_n$ and direct neutron decay asymmetry measurements.  It will be interesting to see if the two methods
agree at that level of precision.

\begin{comment}
In this paper we use the value $g_A = 1.2749(9)$, based on the PDG
value for $\tau_n $ and $V_{ud}$ given in Table~\ref{tab:input}.
 We should note, however, that a recent trapped neutron lifetime experiment at Los Alamos~\cite{Pattie:2017vsj}
 with very small systematic uncertainties finds $\tau_n = 877.7(7)\,{\rm s}$,
 in strong support of earlier trapped neutron results.
 Roughly estimating the effect of the new result on the neutron lifetime average suggests a preliminary
 average $\tau_n^{\rm ave.} = 879.3(9)\,{\rm s}$.
This shorter average lifetime leads to a larger $g_A = 1.2757(7)$ which is
very consistent with the most recent direct neutron decay asymmetry measurements of $g_A$~\cite{Patrignani:2016xqp}.
Of course, a larger $g_A$ used as input
will lead to a larger $\bar{g}_P^{\rm MuCap}=8.24(84)$, but one still fully consistent with theory,
$\gpC^{\rm theory}=8.25(25)$. 
\end{comment}

For now, the value of \raq\ obtained from the $z$ expansion fit to neutrino-nucleon
quasi-elastic scattering together with the MuCap singlet muonic
Hydrogen capture rate $\Lambda_{\rm singlet}^{\rm MuCap}$
can be used in Eq.~(\ref{eq:independent}) to obtain a muon based value, $g_A = 1.276(8)_{r_A^2}(8)_{\rm MuCap} =1.276(11)$.
That overall roughly $\pm 1\%$ sensitivity is to be compared with the current, better than $\pm 0.04\%$,
determination of 
\ga\ from the electron based neutron lifetime that we have been using
in our text.
The good agreement can be viewed as a test of electron-muon universality in semileptonic charged current interactions
at roughly the $1\%$ level by determining
$g_A^e/g_A^{\mu}=0.9996(86)$.  A similar, but not identical
 test is provided by the ratio $( \Lambda_{\rm singlet}^{\rm theory}/ \Lambda_{\rm singlet}^{\rm MuCap})^{1/2}=0.9998(70)$, 
since the theory prediction uses primarely
electron  based parameters. The latter test is closely related to
the ratio of electroweak coupling constants
$g_e/g_\mu=0.9996(12)$  determined from the $\pi\rightarrow e
\nu(\gamma)/\pi\rightarrow\mu \nu(\gamma)$ branching ratio~\cite{Aguilar-Arevalo2015a}.
Leptonic pion decays are considered to be one of the 
best experimental tests of weak charged current e-$\mu$
 universality. Somewhat surprisingly, they are only a factor of 6 better
 than the comparison of theory with experiment in muon capture.
We have described how a factor of 3 improvement in the MuCap capture rate may be experimentally feasible.
A similar factor of 3 or even much better improvement in $r_A^2$ seems possible from lattice QCD first
principles calculations.
Together, such advances would provide a muon based determination of $g_A$ to about $\pm 0.2-0.3\%$
and improve the capture based test of  electron-muon universality at about a factor of 4.
Such a comparison is graphically illustrated in Fig.~\ref{fig:gA}
where the current electron determination of \ga\
from the neutron lifetime is represented by the narrow horizontal band. (A shift in the neutron lifetime
would displace the band up or down.) Muon capture constraints depend on $r_A^2$ and the singlet capture rate as illustrated
by the vertical dashed lines and shaded sloped bands.

\section{Towards a more precise $r_A^2$}
\label{sec:impact}

The momentum dependence of nucleon form factors is critical in
many physical processes.
The various nucleon radii, defined for each form factor analogously to
Eq.~(\ref{eq:FAlowq2}), parameterize this momentum dependence at low $q^2$.
In fact, for momentum transfers $|q^2| \lesssim {\rm few}\;{\rm GeV}^2$,
the form factors become approximately linear functions in the
$z$ expansion: 
  Even in the high-statistics datasets for electromagnetic form factors, the
  curvature and higher order coefficients of the $z$ expansion are
  only marginally different from zero~\cite{Hill:2010yb}.
This emphasizes the prominence of the form factor charges (i.e., normalizations at $q^2=0$)
and radii (i.e., slopes at $q^2=0$).%
\footnote{The $|q^2|\lesssim  {\rm few}\;{\rm GeV}^2$ regime encompasses many physics applications. 
  At larger momentum transfers, inelastic processes compete with the
  elastic process that is determined by the form factors.  For a discussion of $F_A(q^2)$ at
  large $|q^2|$ see Ref.~\cite{Anikin:2016teg}.}
In a nonrelativistic picture, the form factor radii can be interpreted in terms of
nucleon structure.  For example, the electric charge form factor of the proton
represents the Fourier transform of the proton charge distribution, and $r^2$ is readily identified
as a mean-square radius in this picture.
Similarly, the isovector axial form factor can be interpreted
as the Fourier transform of a spin-isospin distribution within the nucleon.
Independent of intuitive nonrelativistic models, the form factor charges and radii
systematically describe the response of nucleons to weak and electromagnetic probes. 

Since the form factors are approximately linear over a broad $q^2$ range, the
radii constrain (and can be probed by) a variety of processes.
For example, the proton charge radius that is probed at $\sim$eV energy
scales using hydrogen spectroscopy can be compared with measurements
at $\sim$GeV energy scales using elastic electron-proton scattering.%
\footnote{
  For a discussion of scheme conventions relevant to this comparison, see
  Ref.~\cite{Hill:2016gdf}.
  }
Similarly, constraints on the axial radius from low-energy
muon capture translate to constraints on higher-energy neutrino scattering
processes.  In Sec.~\ref{sec:impact_rA2} below, we highlight an important
application to quasielastic neutrino scattering and the 
precision neutrino oscillation program. 

The current uncertainty on $r_A^2$ from the $z$ expansion fit to
neutrino-deuteron scattering data is about 50\%.  In this paper, we have shown that the
MuCap experiment already provides similar sensitivity and a future factor of 3 improvement
in a MuCap like experiment could lead to a roughly 20\% determination of $r_A^2$.
In Sec.~\ref{sec:other}, we address the capability of other approaches to the precise determination of $r_A^2$.
As we shall see, currently, it seems that dedicated lattice studies and  neutrino scattering experiments offer the
best opportunities.

\subsection{Impact of improved $r_A^2$ on accelerator neutrino cross sections \label{sec:impact_rA2}}

The discovery of neutrino masses, mixing and oscillations provides our
first real indication of ``new physics", beyond Standard Model
expectations. The source of those effects is likely to arise from
very short-distance phenomena that may require new
technologies and high energy colliders to unveil.  However, in the
meantime, improvements in neutrino oscillation measurements can still
provide important new discoveries.  In that regard, ongoing and proposed neutrino
oscillation experiments will address the following questions: Is CP
violated in neutrino mixing? Are the neutrino masses ordered in
magnitude in the same way or a different way than their charged lepton
counterparts? Do neutrinos have additional interactions with matter
that can be explored through neutrino oscillation interferometry?
Answers to those questions could help explain the source of the
matter-antimatter asymmetry of our universe, a deep fundamental
mystery tied to our very existence.

Neutrino-nucleus interaction cross sections at GeV energies are critical to extracting
fundamental neutrino properties and parameters from long baseline oscillation
experiments~\cite{Adams:2013qkq,Acciarri:2015uup,Diwan:2016gmz}.
Uncertainties in these cross sections arise from the elementary nucleon level
scattering amplitudes, and from data-driven nuclear modeling for detectors
consisting of carbon, water, argon, 
etc.~\cite{Mosel:2016cwa,Katori:2016yel,Alvarez-Ruso:2017oui}.  
A typical oscillation experiment employs a ``near'' detector, close to the production
source of neutrinos, and a ``far'' detector, located at a sufficiently large distance
to allow for observable oscillations.
Naively, the near-far detector comparison can be used to avoid reliance on neutrino
interaction cross sections.   However, a number of effects do not cancel in this comparison: 
flux differences between near and far (e.g. due to oscillation effects and neutrino beam divergence);
flavor dependence of cross sections (e.g. the near detector may constrain $\nu_\mu$ cross sections,
whereas the far detector may search for $\nu_e$ appearance signal); 
and degeneracies between errors in neutrino energy reconstruction from 
undetected particles (such as neutrons, and other sub-threshold particles)
and errors from neutrino interaction uncertainties~\cite{Alvarez-Ruso:2017oui}. 

At the nucleon level, the $q^2$ dependence of the axial form factor is
an important source of uncertainty.  This uncertainty directly impacts
the final cross section, but also complicates the validation of nuclear,
flux, and detector modeling, all of which are predicated on quantitatively
understanding the simplest quasielastic process.  
As an example, the MiniBooNE~\cite{AguilarArevalo:2010zc} analysis of quasielastic neutrino-carbon scattering data
yielded $r_A^2(\rm MiniBooNE)=0.26(7)\;{\rm fm}^2$, in tension with historical values obtained from neutrino-deuteron
scattering data.  Without quantitative control over the nucleon-level amplitudes it
is not possible to unambiguously identify the source of the discrepancy.  

As a proxy for the relevant class of neutrino observables, let us
consider the quasielastic neutrino-neutron cross section at
neutrino energy $E_\nu = 1\,{\rm GeV}$.
Assuming the dipole ansatz for $F_A(q^2)$, with~\cite{Bodek:2007ym} $r_A^2(\rm dipole) = 0.454(13)\,{\rm fm}^2$, 
this cross section may be evaluated as%
\footnote{For definiteness we employ the remaining parameter and form factor choices of Ref.~\cite{Meyer:2016oeg}.}
\begin{linenomath*}
\begin{align}
  \sigma_{\nu n \to \mu p}(E_\nu = 1\,{\rm GeV},\, {\rm dipole}) &= 10.57(14) \times 10^{-39}\,{\rm cm}^2 \,,
\end{align}
\end{linenomath*}
where for the present illustration, we neglect
uncertainties from sources other than $F_A(q^2)$, such as radiative corrections and vector form factors.
Using instead the $z$ expansion representation of $F_A(q^2)$ in Ref.~\cite{Meyer:2016oeg}, the result is 
\begin{linenomath*}
\begin{align}
   \sigma_{\nu n \to \mu p}(E_\nu = 1\,{\rm GeV},\, z\,\, {\rm exp.}) &= 10.1(9) \times 10^{-39}\,{\rm cm}^2 \,, 
\end{align}
\end{linenomath*}
i.e., an uncertainty of order 10\%, an order of magnitude larger than the uncertainty obtained
from the corresponding dipole prediction.

\begin{figure}[h!]
  \begin{center}
  \includegraphics[width=0.7\textwidth]{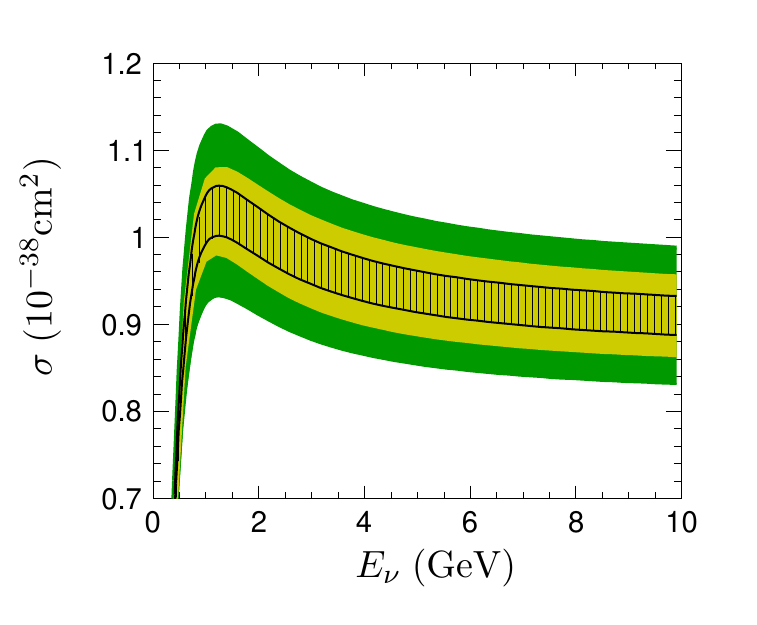}
  \caption{\label{fig:cross}  (color online)
    Quasielastic neutrino-neutron cross section.
    Reference fit of Ref.~\cite{Meyer:2016oeg} in outer solid green band shows the current uncertainty.
    The inner solid yellow band shows the uncertainties independent of $r_A^2$. 
    The hatched black band shows the uncertainty contribution from $r_A^2$, if
    $r_A^2$ would be known to 20\% (using the central value from the reference fit).  In
    that case, the $r_A^2$ contribution would be subdominant in
    the total error (quadratic sum of yellow and black hatched), as illustrated at
    $E_\nu=1\,{\rm GeV}$ in Eq.~(\ref{eq:sigmaerr}). 
  }
  \end{center}
\end{figure}

In order to illustrate the impact of improved constraints on $r_A^2$, we begin by reproducing
the fits of Ref.~\cite{Meyer:2016oeg}, using in addition to the neutrino-deuteron scattering data, an external
constraint on $r_A^2$ (coming, e.g. from muon capture).
The results are displayed in Fig.~\ref{fig:cross}.  Here we first compare the
reference fit to a fit where the slope ($\propto r_A^2$) is constrained to a particular value
(chosen for illustration as the central value $r_A^2=0.46\,{\rm fm}^2$ of the reference fit).
The yellow band in
the figure represents the cross section uncertainty that would result from an external radius
constraint with negligible error.  
As we noted above, $F_A(q^2)$ becomes approximately linear when expressed as a Taylor expansion
in $z$, in the sense that curvature in $z$ and higher order $z$ expansion coefficients, are consistent
with zero, within errors~\cite{Meyer:2016oeg}.  However, these coefficients, when varied over their allowed range,
contribute to the error budget, represented by the width of the yellow band in the figure. 

To illustrate the impact of a finite uncertainty on the external radius constraint, we recompute
the cross section using an external radius constraint that differs by
$\pm 20\%$ from the chosen central value (recall that this corresponds
to the level of precision on $r_A^2$ attainable by
a future 0.33\% muon capture rate measurement).  This variation is represented by the black hatched band
in the figure.  The uncertainties represented by the yellow and hatched bands should be 
added in quadrature to obtain the total cross section error.
At an illustrative $E_\nu=1\,{\rm GeV}$, the result is summarized by 
\begin{linenomath*}
\begin{align}\label{eq:sigmaerr}
  \sigma_{\nu n \to \mu p}(E_\nu = 1\,{\rm GeV},\, z\,\, {\rm exp.}) &= 10.2 \pm {0.47} \pm
  \left( 0.28 {\delta r_A^2/r_A^2 \over 20\%} \right) \times 10^{-39}\,{\rm cm}^2 \,. 
\end{align}
\end{linenomath*}
External constraints on $r_A^2$, used in conjunction with the existing deuteron target
neutrino scattering data, 
can thus lead to a halving of the uncertainty on the elementary
signal cross section for long baseline neutrino experiments.
Advances in our quantitative understanding of neutrino scattering,
through improvements in $r_A^2$, heavy nuclear target modeling
and direct precise neutrino cross-section measurements
will allow us to fully exploit the planned sensitivity of future oscillation experiments.

\subsection{Other constraints and applications \label{sec:other}}

Given the importance of $r_A^2$, and more generally $F_A(q^2)$, let us understand
what complementary information exists from other approaches.  This information
comes from theoretical approaches to determine $F_A(q^2)$ from the QCD Lagrangian;
and from experimental measurements using weak and electromagnetic probes of the nucleon. 

\subsubsection{Lattice QCD \label{sec:lattice}}

Lattice QCD is a computational method for determining low energy properties of
hadrons based on first principles starting from the QCD Lagrangian.%
\footnote{For a brief introduction and references see the lattice QCD review of 
   S.~Hashimoto, J.~Laiho and S.~R.~Sharpe
   in Ref.~\cite{Patrignani:2016xqp}.}
This method has reached a mature state for meson properties.%
\footnote{For a review and further references, see Ref.~\cite{Aoki:2016frl}.}
Nucleons present an additional challenge for lattice simulations,
owing to a well-known noise problem~\cite{DeGrand:1990ss}.
A variety of approaches are being taken to explore and address the simultaneous
challenges of excited states, lattice size, finite volume, as well as
statistical noise.
In many cases, the need to extrapolate from unphysically large light
quark masses is overcome by performing the lattice calculation at (or near)
the physical masses.
Background field and correlator derivative techniques are being explored to
optimize the isolation of nucleon properties.%
\footnote{For recent examples and further references,
  see Refs.~\cite{Tiburzi:2017iux,Savage:2016kon,Bouchard:2016heu,Berkowitz:2017gql,Alexandrou:2016hiy}.}

Recent computations of the isovector axial charge with a complete stated
error budget include:
$g_A = 1.195(20)(33)$~\cite{Bhattacharya:2016zcn},
where the first error is due to extrapolating
in lattice spacing, lattice volume and light quark masses, and 
the second error is statistical and other systematics; 
and
$g_A = 1.278(21)(26)$~\cite{Berkowitz:2017gql},
where the first error is statistical and fitting systematics, and
the second error is due to model selection in the chiral and continuum
extrapolation. 
Other recent preliminary results and discussions
may be found in
Refs.~\cite{Abramczyk:2016ziv,Yamazaki:2015vjn,Liang:2016fgy,Abdel-Rehim:2015owa,Djukanovic:2016ocj,Meyer:2016kwb,Green:2017keo,Alexandrou:2017hac}.
We remark that QED radiative corrections are below the current lattice QCD sensitivity, and the
details of the $g_A$ definition in the presence of radiative corrections are thus not yet relevant for this comparison.
Note also that the isovector quark current is scale independent in the usual $\overline{\rm MS}$ scheme used to present
lattice results. 

Lattice QCD is approaching the few percent level for $g_A$.
A complete calculation of $r_A^2$ that would rival the precision of
neutrino-nucleon scattering and muon capture is not yet available from lattice QCD.
However, 
illustrative values have been obtained, typically using simplified functional forms
for the $q^2$ behavior, unphysically large light quark masses, and/or neglect
of strange and charm quarks.
A dipole form factor ansatz fit to two-flavor lattice QCD extractions of
$F_A(q^2)$~\cite{Alexandrou:2017hac} found a result, $r_A^2=0.266(17)(7)\,{\rm fm}^2$,
where the first error is statistical and the second is systematic due to excited states;
this result lies closer to the ``large $m_A$'' MiniBooNE dipole result~\cite{AguilarArevalo:2010zc}
than to the ``small $m_A$'' historical dipole average~\cite{Bernard:2001rs,Bodek:2007ym}. 
A $z$ expansion fit to $F_A(q^2)$ obtained using three-flavor QCD with physical strange quark mass,
and heavier-than-physical up and down quark masses (corresponding to pion mass 317~MeV)~\cite{Green:2017keo},
yielded $r_A^2=0.213(6)(13)(3)(0)\,{\rm fm}^2$, where the uncertainties are from
statistics, excited states, fitting and renormalization.
A first order $z$ expansion fit to $F_A(q^2)$ using two-flavor QCD,
extrapolated to physical pion mass~\cite{Capitani:2017qpc} yielded
$r_A^2=0.360(36)^{+80}_{-88}\,{\rm fm}^2$, where the first error is statistical and the
second error is systematic.
Finally, a $z$ expansion fit to four-flavor lattice
QCD data using a range of lattice parameters~\cite{Gupta2017} yielded
$r_A^2=0.24(6)\,{\rm fm}^2$. 
Some of these $r_A^2$ values are well below the historical dipole value and even disagree somewhat
with our conservative average of $\raq(\rm avg.) =  \raqAN\; \fmq$ in Eq.~(\ref{eq:average}). 
This situation suggests that either remaining lattice corrections
will involve large corrections that will significantly shift the
lattice determination of the radius, 
or perhaps more exciting that a disagreement may
  persist as further lattice progress is made, leading to a new paradigm
  in our understanding of $r_A^2$.  However, at this point, further work is
  needed to obtain precise lattice results with more complete error budgets.

\subsubsection{Pion electroproduction}

Fits to pion electroproduction data have historically contributed to the determination
of the axial radius, with a small quoted uncertainty that can be traced to the assumed
dipole form factor constraint.
The statistical power of available data would be comparable to the neutrino-deuteron
scattering determination, but relies on extrapolations beyond the regime of low
energies where chiral corrections are controlled.
The axial form factor appears in a low energy theorem for
the $S$-wave electric dipole amplitude of threshold charged pion electroproduction
($e^- p \to e^- n \pi^+$)~\cite{Nambu:1997wa,Nambu:1997wb},
\begin{linenomath*}
\begin{align}
  E_{0+}^{(-)}\big|_{m_\pi=0} &= \sqrt{1-{q^2\over 4 m_N^2}}
  {e g_A \over 8\pi f_\pi} \bigg[ F_A(q^2) + {q^2\over 4 m_N^2 - 2q^2} F_M(q^2) \bigg] \,. 
\end{align}
\end{linenomath*}
This low energy theorem is strictly valid in the chiral limit ($m_\pi=0$) for
threshold production (invariant mass $W=m_N+m_\pi$ in final state hadronic system).
The chiral and threshold limits do not commute, but corrections to the low energy theorem
may be calculated within \hb~\cite{Bernard:1992ys}. 
Two complications enter.  First, experimental measurement is difficult, involving the
detection of either a recoiling neutron or a low energy pion.  
Most of the statistical power of available
data involves energies and momentum transfers outside of the regime where
a chiral expansion is reliable.
The data have been interpreted in terms of a phenomenological framework, whose
associated systematic uncertainty is difficult to assess.  
Second, taking at face value the phenomenological extraction of $F_A(q^2)$ at certain kinematic points
from the experimental data, the interpretation as a measurement of the radius
has assumed a dipole shape that strongly influences the result. 

Using the extracted form factor values at particular kinematic points from
Refs.~\cite{Amaldi:1972vf,Brauel:1973cw,DelGuerra:1975uiy,DelGuerra:1976uj,Esaulov:1978ed},
but replacing dipole with $z$ expansion,
Ref.~\cite{Bhattacharya:2011ah} obtained $r_A^2 = 0.55 (17)\,{\rm fm}^2$, compared to
the dipole analysis of Ref.~\cite{Bernard:2001rs} which gave $r_A^2= 0.467 (18) \,{\rm fm}^2$.  
The datasets were selected to coincide with those
that appear in the compilation~\cite{Bernard:2001rs} in order to make a direct comparison with their dipole
fit (cf. Figure~1 of that reference). 
These datasets explicitly list inferred values of $F_A(q^2)$ (see also 
\cite{Amaldi:1970tg,Bloom:1973fn,Joos:1976ng,Choi:1993vt,Liesenfeld:1999mv}). 
Reference~\cite{Liesenfeld:1999mv} provides a value $r_A^2 = 0.449(28) \,{\rm fm}^2$ based
on data at $W=1125\,{\rm MeV}$ and $Q^2=0.117$, $0.195$ and $0.273\,{\rm GeV}^2$, and 
a phenomenological Lagrangian analysis.%
\footnote{This result is obtained from the dipole axial mass
          $m_A=1.077(39)$ GeV, after applying
  the chiral correction $\delta r_A^2 = 0.046\,{\rm fm}^2$~\cite{Bernard:2001rs}.}
Reference~\cite{Friscic:2016tbx} presents data at $W=1094\,{\rm MeV}$ and
$Q^2=0.078\,{\rm GeV}^2$.   
Regardless of the precise choice of dataset,
the error is significantly larger when the strict dipole assumption is relaxed, even when
systematics associated with extrapolations outside of the chiral Lagrangian framework are neglected. 
Further effort is needed before pion electroproduction provides a robust answer for
$r_A^2$.

\subsubsection{Lepton scattering} 

Since the most direct constraints on $F_A(q^2)$ come from neutrino scattering data,
it is natural to ask whether improved measurements are feasible.
The world dataset for neutrino deuteron scattering consists of a few thousand
quasielastic events from bubble chamber data of the 1970s and 1980s.  Systematic
uncertainties from hand-scanning of photographs and from nuclear modeling are comparable
to statistical errors, contributing to the total quoted uncertainty on \raq\ of
$0.22\,{\rm fm}^2$ in Ref.~\cite{Meyer:2016oeg}.  Note that the flux is
determined self consistently from the quasielastic events, so that flux errors
associated with neutrino production are not relevant. 
Although nuclear corrections for deuteron targets are relatively small compared to
heavier nuclei, errors are difficult to quantify at the desired few percent level,
for accelerator neutrino beams of GeV energies.  
Antineutrino data on hydrogen would eliminate even these relatively small corrections.  
Existing antineutrino quasielastic data is very sparse, owing to the combined
penalties of smaller production cross section for creating antineutrino versus neutrino
beams, and smaller scattering cross section for antineutrinos versus neutrinos.
Thus most data were taken in neutrino mode versus antineutrino mode. 
Reference~\cite{Fanourakis:1980si} reported $13\pm 6$ events.
References~\cite{Ahrens:1986xe,Ahrens:1988rr} reported results for
antineutrino-proton scattering inferred from data taken on nuclear (carbon) targets. 
Currently available analysis techniques with an active target detector 
should reduce or eliminate scanning and efficiency systematic corrections.
Modern neutrino beams have much higher flux compared to the beams used for
the existing datasets which would enable either a much smaller detector or
a much larger dataset over a given timescale.
Technical, cost and safety considerations must be addressed in order to
make such a new measurement feasible.

The capture process $\mu^- p \to \nu_\mu n$ in muonic hydrogen, and the
time reversed process $\nu_\mu n \to \mu^- p$ measured in neutrino scattering,
both probe the charged-current component of the isovector axial vector nucleon
matrix element.  By isospin symmetry, this isovector matrix element can also be
accessed via the neutral component.  Parity violating electron-nucleon elastic
scattering~\cite{Beise:2004py,Androic:2009aa}, induced by weak $Z^0$ exchange,
is a probe of this matrix element, but simultaneously involves also isoscalar
and strange quark contributions that must be independently constrained.  
Available data do not have discriminating power to reliably extract
axial radius or form factor shape information. 
For example, the G0 experiment~\cite{Androic:2009aa} analyzed electron-proton and
electron-deuteron scattering data to perform a simultaneous fit of
the isovector axial form factor, and the strange vector form factors,
taking the remaining form factors from other sources.  An
amplitude was measured for $F_A(q^2)$ at $Q^2=-q^2=0.22$ and 0.63~GeV$^2$, but
with insufficient precision to extract shape information. 
The process $e^+ d \to \bar{\nu}_e pp$ is another possibility to access
the charged current nucleon interaction, $e^+ n \to \bar{\nu}_e p$
using electron (positron) beams.
No measurements of this process currently exist.

\subsubsection{Summary of complementary constraints}

\begin{figure}[h!]
  \begin{center}
 \includegraphics[width=0.75\textwidth]{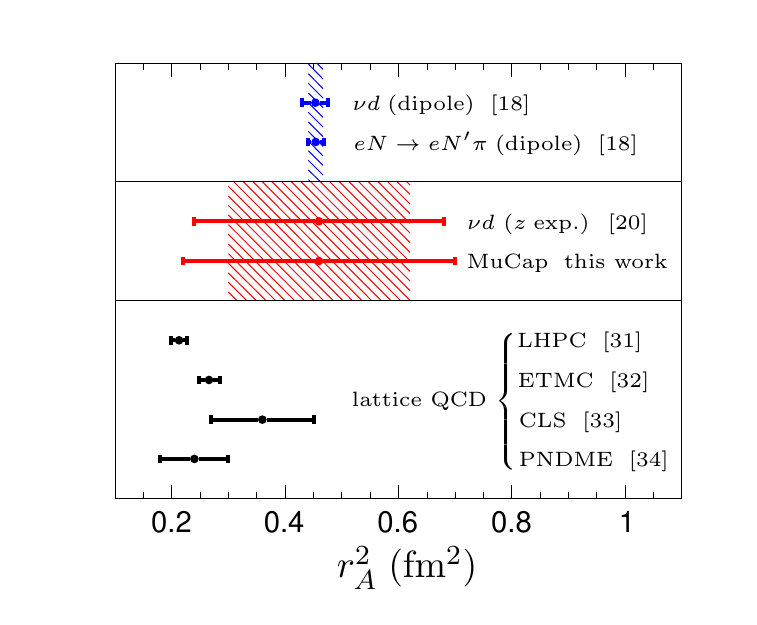}\vspace{-5mm}
  \caption{\label{fig:rA2}
    (color online)
    Squared axial radius determined by different processes, cf. Table~\ref{tab:rA2vals}.
    Top: Dipole fits to $\nu d$ and $eN \to eN^\prime \pi$ data employed before 2016 resulted
    in an average \raq\ with small uncertainty (hatched blue
    band)~\cite{Bodek:2007ym}.
    Middle: Replacing the unjustified dipole assumption with the $z$ expansion allowed a model independent
    extraction of \raq\ from $\nu d$ data and increased the uncertainty. The hashed red region represents
    the best average from this work combining the new determination from MuCap with the $\nu d$ result
    [cf. Eq.~(\ref{eq:average})].
    Bottom: Early lattice QCD results denoted
    by their collaboration acronyms. Several calculations
    tend towards lower values of \raq\ compared
    to the historical dipole average.
  }
  \end{center}
\end{figure}

A range of processes and techniques have potential to help constrain
the nucleon axial radius.   Some of these, such as pion electroproduction and 
parity violating electron-proton scattering, access the form factor and radius indirectly
and suffer significant model-dependent corrections that need to be further
addressed to achieve $\sim 10\%$ accuracy on $r_A^2$.  Lattice QCD and
elementary target neutrino scattering are potentially pristine theoretical or experimental
approaches.  However, lattice QCD has not yet achieved the requisite accuracy,
and hydrogen or deuterium active target neutrino experiments are fraught with
surmountable but difficult technical and safety issues.
Figure~\ref{fig:rA2} displays the range of values for $r_A^2$ as tabulated in
Table~\ref{tab:rA2vals}, including the MuCap determination presented in this paper.
Our average, Eq.~(\ref{eq:average}), is obtained from the $z$ expansion $\nu d$
and MuCap results, which have complete error budgets.  
The future is sure to witness an interesting complementarity between
different approaches to axial nucleon structure, with a wide range of constraints
and applications.

\section{Summary and outlook \label{sec:discussion}}

In this paper we considered the status and prospects of constraints
on the isovector axial nucleon form factor, \FA(\qq),   
which describes a range of lepton-nucleon reactions.
We focused in particular on its 
prominent role in neutrino-nucleus scattering cross sections
underlying neutrino oscillation experiments at accelerator energies.  
A precise knowledge of these cross
sections, for momentum transfers  $|q^2| \lesssim
{\rm few}\;{\rm GeV}^2$, is required
for the next generation of precision
studies of neutrino properties in long baseline oscillation
experiments.
Fully utilizing the oscillation data will require better knowledge of the
structure of \FA(\qq), in concert with data-driven
improvements in heavy nuclear target modeling.  

For many processes, the first two terms in the $q^2$ expansion of 
\FA(\qq)\ can be shown to dominate. 
These terms are parameterized by
$\ga \equiv \FA(0)$,  and \raq, which is proportional to the slope of \FA(\qq) at $\rm \qq
\rightarrow 0$. The axial nucleon coupling, \ga, is precisely determined
from neutron beta decay; we use \ga=\gaN\  in this work.
The nucleon axial radius squared, \raq, was considered
well-determined and uncontroversial for a long time, with $r_A^2(\rm dipole) =
0.454(13)\,\fmq$, derived from a dipole fit to neutrino scattering and pion
electro-production data.  A recent analysis, however, eliminated the 
dipole shape constraint for \FA(\qq), as not justifiable from first
principles.  Using instead the $z$ expansion as a model
independent formalism to enforce properties inherited from the underlying
QCD structure, 
a value \raq($z$ exp.,\;$\nu$)=\raqN $\,\rm fm^2$  was
derived~\cite{Meyer:2016kwb} from a fit to the $\nu d$ scattering data.
The more conservative, but better justified error is an
order of magnitude larger than that from the dipole fit, with nearly
50\% uncertainty.
In this work we assessed some ramifications of this new development, and
in particular reviewed and suggested opportunities to reduce the uncertainty in \raq. 

 We started from the vantage point of muon capture, in particular muon
 capture in the theoretically pristine atom of muonic hydrogen,
 $\mu$H. Muon capture is a charged-current reaction with a small
 momentum transfer, $\qqC\approx -0.9\, m_\mu^2$, so
that \FA(\qqC) is only $\sim 2\%$ smaller than \FA(0).  In the past, the uncertainty introduced
by the error in $r_A^2(\rm dipole)$ was considered negligible, and capture
in $\mu$H was used to determine the nucleon pseudoscalar coupling \gpC.
The  recent 1\% MuCap measurement of the spin singlet muonic hydrogen capture 
rate, $\Lambda_{\rm singlet}^{\rm MuCap} =715.6(7.4)\,\si$, determined
$\bar{g}_P^{\rm MuCap}= 8.06(55)$, using 
theory and form factors available at the time. The agreement of this result with the
precise prediction
of \hb\ is considered an important test of the chiral structure of QCD.
Given the dramatically increased uncertainty in \raq, we addressed the
following questions, answering both in the affirmative: 
Does the comparison of \gpC\ between experiment and theory still provide
a robust test of \hb? 
And, in a reversal of strategy, can muon capture be used to determine
a competitive value of \raq?

High precision is required both in theory and experiment to utilize the
small effect of \raq\ on muon capture.  In this paper, we have reduced the uncertainty in the
electroweak radiative corrections to muon capture to the 0.10\% level and extracted an updated
value, $\bar{g}_P^{\rm MuCap} = \gpMuCapN$, from the MuCap experiment.
Agreement with the updated theoretical prediction from \hb, $\bar{g}_P^{\rm theory} = \gpCN$,
remains excellent.  
It confirms expectations at a sensitivity level of $\pm 8\%$, 
weakened only slightly (from $\pm 6\%$) by the larger uncertainty in \raq($z$ exp.,\;$\nu$)
compared to \raq(dipole). 
The MuCap result was also used to provide a self-consistent test of the pion-nucleon
coupling and to obtain a roughly $\pm 1\%$ muon-based value of $g_A$,
which was found to be in agreement with the electron-based value traditionally
extracted from neutron decay (thereby, testing electron-muon
universality). Of course, all such tests would be improved by a better
independent determination of $r_A^2$ and the factor of 3 improvement in a
next generation muon capture experiment advocated here.

As a novel application of the muonic Hydrogen capture rate, we
explored its use as an
 alternative method for determining the nucleon axial radius squared.
 Using the rather precise theoretical expression 
 for $\bar{g}_P^{\rm theory}$ from \hb,
 along with
updated radiative corrections 
and form factors as input, we found
$r_A^2(\mu{\rm H}) = \raqMuCapN \;{\rm fm}^2$ from the MuCap singlet capture rate measurement.
Combining that finding with the $z$ expansion neutrino-nucleon 
scattering result led to a weighted average $r_A^2({\rm ave.})
=\raqAN {\rm fm}^2$.  
We also examined the possibility of improving the MuCap experiment by
roughly a factor of 3 and thereby 
determining $r_A^2$ to about $\pm 20\%$. 
As demonstrated, that level of accuracy would be sufficient to reduce the \raq\
induced uncertainties in neutrino-nucleon scattering to a subdominant
level. Moreover, it would start to become a standard for comparison with other methods of
$r_A^2$ determination, several of which were discussed.  For such comparisons, lattice
gauge theory calculations appear to hold the most promise.  Although that Monte Carlo approach to QCD is
still not fully mature as applied to $r_A^2$, 
it promises a first principles strong coupling method that in time
should reach high precision.
Some early current efforts seem to suggest a significantly smaller value
of $r_A^2$ compared to historical dipole averages, but it is still too early to 
scrutinize or average the lattice results in a meaningful way.  Future confrontation
between experiment and lattice QCD will be interesting to watch and could provide
surprises. 

The nucleon axial radius has reached an exciting new stage.
Until recently, it was thought to be well determined 
by dipole form factor fits to neutrino-nucleon scattering and
electroproduction measurements.  However, driven especially by the need 
for better neutrino cross section predictions, that common lore has
been replaced by more conservative healthy skepticism. 
The axial vector form factor is now being approached from many directions,
with the potential to challenge conventional dogma as it enters
a new precision era.

\vskip 0.2in
\noindent
{\bf Acknowledgments}
\vskip 0.1in
\noindent
We thank M.~Hoferichter for helpful discussion of the pion-nucleon coupling. 
We acknowledge the Institute for Nuclear
Theory at the University of Washington, where the idea for
this review was conceived.
R.~J.~H. thanks TRIUMF for hospitality where a part of this work was performed.
Research of R.~J.~H. was supported by a NIST Precision Measurement Grant. 
Research at Perimeter Institute is supported by the Government of Canada through the Department of Innovation,
Science and Economic Development and by the Province of Ontario through the Ministry of Research and Innovation.
Fermilab is operated by Fermi Research Alliance, LLC under Contract No. DE-AC02-07CH11359 with the United States
Department of Energy.
The work of P.~K. was supported by the U.S. Department of Energy Office of Science,
Office of Nuclear Physics under Award Number DE-FG02-97ER41020.
The work of W.~J.~M. was supported by the U.S. Department of Energy under grant DE-SC0012704.
The work of A.~S. was supported in part by the National Science Foundation under Grant PHY-1620039.

\newpage

\bibliography{NucleonAxialRadius}

\end{document}